\documentclass[11pt]{article}

\usepackage[numbers]{natbib}
\usepackage[ruled,vlined]{algorithm2e}
 
\usepackage{amsmath, amssymb, amsthm}
\usepackage{amsfonts}
\usepackage{graphicx}
\usepackage{pgfplots}
\usepackage{multirow}
\usepackage{indentfirst}
\usepackage{pifont}
\usepackage{bbm}

\usepackage[toc,page]{appendix}
\usepackage[font=small,labelfont=bf]{caption}

\usepackage[unicode=true, pdfusetitle, bookmarks=true, bookmarksnumbered=true, bookmarksopen=true, bookmarksopenlevel=1, breaklinks=true, pdfborder={0 0 1},
backref=false, colorlinks=true]{hyperref}
\hypersetup{
    linkcolor=blue,
    citecolor=blue,
    urlcolor=blue,
    filecolor=blue,
    pdfpagelayout=OneColumn,
    pdfnewwindow=true,
    pdfstartview=XYZ,
    plainpages=false
}
\usepackage{authblk}

\newcommand{\cmark}{\ding{51}}
\newcommand{\xmark}{\ding{55}}
\pgfplotsset{compat=1.17}

\newcommand{\RR}{\mathbb{R}}
\newcommand{\FF}{\mathcal{F}} 
\newcommand{\Tau}{\mathcal{T}}

\renewcommand{\leq}{\leqslant}
\renewcommand{\geq}{\geqslant}
\renewcommand{\epsilon}{\varepsilon}

\DeclareMathOperator*{\esssup}{ess\,sup}

\newcommand{\D}[1]{\ \mathrm{d}#1}
\renewcommand{\-}{\,\scalebox{0.75}[1.0]{-}}
\DeclareMathOperator{\Vvar}{Var}
\newcommand{\Var}[1]{\Vvar\left(#1\right)}
\newcommand{\EV}[1]{\mathbb{E}\!\left[#1\right]}
\newcommand{\indic}{\mathbbm{1}}
\newtheorem{remark}{Remark}[section]
\newtheorem{proposition}{Proposition}[section]

\begin{document}

\title{Simultaneous upper and lower bounds of American-style option prices with hedging via neural networks}
       
\author[1,2]{Ivan Guo\thanks{Ivan Guo's work was partially supported by the Australian Research Council (Grant DP220103106) and CSIRO Data61 Risklab.}}
\author[3]{Nicolas Langren\'e\thanks{Nicolas Langren\'e's work was supported in part by the Guangdong Provincial Key Laboratory of Interdisciplinary Research and Application for Data Science, BNU-HKBU United International College, project code 2022B1212010006, and in part by the UIC Start-up Research Fund UICR0700041-22.}}
\author[1]{Jiahao Wu}

\affil[1]{\small School of Mathematical Sciences, Monash University, Melbourne, Australia}
\affil[2]{\small Centre for Quantitative Finance and Investment Strategies, Monash University, Australia}
\affil[3]{\small Guangdong Provincial Key Laboratory of Interdisciplinary Research and Application for Data Science, BNU-HKBU United International College, Zhuhai, China}

\date{\today}
\maketitle
\begin{abstract}
In this paper, we introduce two novel methods to solve the American-style option pricing problem and its dual form at the same time using neural networks. Without applying nested Monte Carlo, the first method uses a series of neural networks to simultaneously compute both the lower and upper bounds of the option price, and the second one accomplishes the same goal with one global network. The avoidance of extra simulations and the use of neural networks significantly reduce the computational complexity and allow us to price Bermudan options with frequent exercise opportunities in high dimensions, as illustrated by the provided numerical experiments. As a by-product, these methods also derive a hedging strategy for the option, which can also be used as a control variate for variance reduction. 
\end{abstract}
\section{Introduction}
Pricing American-style options is a type of optimal control/stopping problem for which numerical methods have been extensively explored due to the lack of analytical solutions. However, classical methods based on partial differential equations and binomial trees become expensive computationally when there are multiple factors impacting the value of the option, a limitation known as the \textit{curse of dimensionality}. To circumvent this difficulty, simulation-based methods have been extensively explored \citep{tilley1993, barraquand1995, carriere1996, longstaff2001, vanroy2001, bouchard2004discrete, broadie2004, bally2005quantization, kolodko2006iterative, bouchard2012monte, ludkovski2018kriging}. By directly solving the pricing problem, these methods typically generate a candidate optimal stopping strategy and a lower bound on the price, which is more in the interest of the buying party. On the other hand, option sellers would be more interested in an upper bound. \citet{kogan2004} and \citet{rogers2002} independently explored the duality of the pricing problem, based on which a variety of methods have been proposed \citep{andersen2004primal, jamshidian2004numeraire, belomestny2009true, rogers2010dual, schoenmakers2013optimal} to derive an upper bound on the option price by solving its dual problem.

Among the dynamic programming-based methods, the Least Squares Monte Carlo (LSMC) method \citep{longstaff2001, vanroy2001} has gained much popularity. In search of the optimal stopping strategy, continuation values are approximated by a pre-defined, static basis via linear regression. However, as the dimension of the problem increases, the number of basis functions significantly increases and the method can become numerically unstable. 
Various studies, including \citet{kohler2010pricing}, \citet{lapeyre2021} and \citet{herrera2021optimal}, have proposed to replace the linear regression in the LSMC method by neural networks (NNs). 
Additionally, \citet{goudenege2020machine} and \citet{goudenege2021variance} have employed Gaussian process regression to estimate the continuation value. \citet{reppen2022neural} applied NNs to parameterise the stopping boundary. Moreover, \citet{bayer2021randomized} have devised a forward and a backward algorithm to approximate the stopping strategy by randomising them with independent noises, while \citet{gonon2024deep} has utilised neural networks to directly approximate the value function and showed that the method is free of the curse of dimensionality. Other works \citep{han2018solving, raissi2024forward, chen2021deep, germain2021neural, na2023efficient, gao2023convergence} have explored the application of deep learning in option pricing by addressing the corresponding partial differential equations (PDEs) or backward stochastic differential equations (BSDEs).

Besides option pricing, hedging strategies are crucial in risk management. Most existing methods for generating hedging strategies either involve taking the first derivative of approximated option value functions \citep{bally2005quantization, bouchard2012monte, jain2015stochastic, na2023efficient} or approximating the function representing the difference between option values at different times once the option has been priced \citep{becker2019deep, beck2020overview}. However, the efficiency of these strategies relies on the accurate differentiation of the estimated continuation value function. Since functions with similar values can have very different derivatives, even satisfying approximations of the value process can lead to ineffective hedging strategies. 

 The primary contribution of our work lies in incorporating the dual formulation of the option price into the modified LSMC method to design algorithms that concurrently produce both lower and upper bounds of the option price. Moreover, our method facilitates the derivation of hedging strategies as an immediate by-product, computed directly from the dual martingale used in the upper bound estimate instead of the differentiation. Unlike traditional methods, our approach offers hedging strategies at all times before maturity, not just at exercise times, and can serve as a control variate to reduce variance, thereby yielding a more accurate lower bound.
\citet{becker2020pricing} proposed a method to price Bermudan options in high-dimensions. However, in their method, they first find a stopping strategy to approximate a lower bound, based on which they then derive an upper bound using nested Monte Carlo. Similarly, their hedging strategy is also based on the stopping strategy with another independent simulation. Their other work \citep{becker2019deep} has a similar structure but approximates the stopping strategy instead. In the case of pricing Bermudan options with frequent exercise opportunities, which approximates an American-style option, the computational cost can be very high as the cost of nested simulation increases quadratically with the number of stopping opportunities. Similar methods designed by \citet{lokeshwar2022explainable, belomestny2009true} do not require nested simulations, but the derivation of a biased upper estimate is separate from the determination of the stopping strategy. The work by \citet{hure2021convergence} on reflected BSDEs resolution shares some resemblance, but it only generates a point estimate, and the details in the dynamic programming are different.
 
In addition, we present the use of one global network instead of a series of networks in the derivation by treating time as an additional state variable. Global networks have been introduced to solve semi-linear PDEs \citep{chan2019machine} and other control problems \citep{gobet2005sensitivity, fecamp2019risk}. In such stopping problems, the target values are known when the training starts as they are outputs of the problem, rather than inputs. However, the training targets are unavailable at the outset of the problem. We propose to alternate the update of stopping strategies and the network training till it produces satisfactory results.

This paper is structured in the following order. Section 2 lays out the theoretical groundwork for combining the LSMC algorithm with the dual formulation. In Section 3, we introduce the numerical methods devised and then present various variants in Section 4. Section 5 is dedicated to demonstrating numerical results in both low- and high-dimensional settings, and then we conclude in Section 6.
 
\section{Problem formulation}

Consider an American option with maturity $T>0$. Let $(\Omega, \mathcal{F},\mathbb{F}=(\mathcal{F}_t)_{t\in[0,T]}, \mathbb{Q})$ be a filtered probability space, where $\mathbb{F}$ is the augmented filtration of a $d$-dimensional Brownian motion $(W_{t})_{t\in[0,T]}$, and $\mathbb{Q}$ is the equivalent martingale measure. 

Define $\beta_t=e^{rt}$ as the value of the risk-free account at $t\in[0,T]$, where the constant $r$ is the risk-free interest rate. The price of the option is based on $d$ risky assets whose value process $(S_{t})_{t\in[0,T]}$ is Markovian and is the solution to the SDE $$dS_t= rS_tdt+\sigma(t, S_t)  dW_t,$$ where 
$\sigma \colon [0,T] \times \RR^{d} \to \RR^{d\times d}$ is assumed to satisfy sufficient regularity conditions to ensure the well-posedness of the equation. 

\subsection{The lower bound of the option price}

Let $(Z_{t})_{t\in[0,T]}$ denote the $\mathbb{F}$-adapted continuous discounted payoff process of the option satisfying $\mathbb{E}[\sup_{t\in [0,T]}Z_t]<\infty$. Let $\tau \colon \Omega \to [0, T]$ be a stopping time, and $\mathcal{T}$ be the set of all stopping times with respect to the filtration $\mathbb{F}$. Then, the value of the American option at time $t$ discounted back to time $0$ is 
$$V_{t} = \esssup_{\tau\in\mathcal{T}, \tau \geq t} \EV{Z_\tau\Big|\mathcal{F}_{t}},$$ 
and in particular the value at time zero is 
$$V_{0} = \esssup_{\tau \in \mathcal{T}} \EV{Z_{\tau}}.$$ 

For any specific stopping strategy $\tau'\in\mathcal{T}$, we have $V_{0}' = \EV{Z_{\tau'}} \leq \underset{\tau\in \mathcal{T}}{\esssup}\,\EV{Z_\tau}=V_0.$
Hence the estimate of an American option price given by one strategy is a lower bound of the real value.

\subsection{The upper bound of the option price}
Denote by $\mathcal{M}^{U\hspace{-0.06em}I}$ the set of all uniformly integrable martingales with  initial state set to zero. Since the discounted option value process $(V_{t})_{t\in[0,T]}$ is a supermartingale of class D, it has a unique Doob-Meyer decomposition:
\begin{equation}\label{eq: DoobMeyer}
V_{t} = V_{0} + M^{*}_{t} - A^{*}_{t},
\end{equation}
where $M^{*} \in \mathcal{M}^{U\hspace{-0.06em}I}$, and $A^{*}$ is a predictable non-decreasing process with $A^{*}_{0} = 0$.

\noindent The American option pricing problem has a dual form:  
\begin{equation}\label{eq: Duality}
    V_0=\inf_{M\in\mathcal{M}^{U\hspace{-0.06em}I}} \EV{\sup_{t\in[0,T]} Z_{t}-M_{t}},
\end{equation}  
and the infimum is attained at $M = M^{*}$. We refer to \citet{ rogers2002, kogan2004} for proofs of this duality.

Denote $\mathcal{M}\subset\mathcal{M}^{U\hspace{-0.06em}I}$ as the set of martingales that are both uniformly integrable and square integrable. We restrict our search for $M^*$ within the set $\mathcal{M}$. This does not pose a problem in our numerical experiments as the optimal martingales corresponding to the options we price satisfy this condition. Since $M^*\in\mathcal{M}$ and is adapted to the Brownian filtration $\mathbb{F}$, the Brownian martingale representation theorem states that there exists a predictable process $H$ with values in $\RR$ such that $\EV{\int_{0}^{T} H^{2}_{s} \D s} < \infty$, and 
\begin{equation}\label{eq: MRepresentation}
M_{t}^{*} = M_{0}^{*} + \int_{0}^{t} H_{s} \D W_{s}.
\end{equation}
This allows us to estimate the optimal martingale $M^*$ by approximating the process $H$ numerically, and then generate an upper bound of the option price.

\subsection{The hedging strategy}
Consider a measurable adapted process $(J_{t})_{t\in[0,T]}$ with values in $\mathbb{R}^{d+1}$, where $J^{i}$ is the number of units of the $i$-th asset held in a portfolio consisting of $d$ risky assets and one risk-free asset. The value of the portfolio at time $t$ is 
$$U_{t} = J_{t}^{0}\beta_{t}+\sum_{i=1}^{d}J_{t}^{i}S_{t}^{i}.$$
The process satisfies the condition 
$\int_{0}^{T}|J_{u}|^{2}\D u 
= \sum_{i=1}^{d} \int_{0}^{T} |J_{u}^{i}|^{2} \D u < \infty$ a.s, and it is a self-financing hedging strategy if
\begin{equation}\label{eq: self-financing}
    \frac{U_{t}}{\beta_{t}} = U_{0}+\int_{0}^{t} J_{u}\beta_{u}^{-1}\sigma(u, S_{u}) \D W_{u}.
\end{equation}

Combining the Doob-Meyer decomposition (\ref{eq: DoobMeyer}) and the Brownian martingale representation (\ref{eq: MRepresentation}), we obtain 
\begin{equation}\label{eq: DBandMR}
V_{t} = V_{0} + \int_{0}^{t} H_{u} \D W_{u} - A_{t}^{*}. 
\end{equation}
For the portfolio to super-replicate the option, we need $\frac{U_{t}}{\beta_{t}} \geq Z_{t}$ for all $t\in[0,T]$. It is well-known that the cheapest such portfolio satisfies $U_0=V_0$ and $\frac{U_{t}}{\beta_{t}}\geq V_{t}\geq Z_{t}$ for all $t\in[0,T]$. Comparing equations (\ref{eq: self-financing}) and (\ref{eq: DBandMR}), we see that this can be achieved by setting
\begin{equation*}
  J_{t} = \frac{\beta_{t}H_{t}}{\sigma(t, S_t)} .  
\end{equation*}
Hence, the hedging strategy $J$ can be computed directly from the process $H$.
The process $A^{*}$ can be interpreted as the losses incurred every time the optimal exercise opportunity is missed. 


\section{Valuing an American option numerically}
From now on, we only approximate American options by Bermudan options whose exercise times are restricted to the discrete set 
$t_{i} = t_{0} + i\cdot \Delta t$, for $i \in \{1,...,n\},$ where $\Delta t=\frac{T}{n}$. Note that since the pricing progress proceeds backward in time, in this paper, at $t_{i}$, the previous step refers to $t_{i+1}$ and the next step refers to $t_{i-1}$. 

We design two algorithms based on the combination of resolutions of both the primal and the dual problem. One uses a series of neural networks, and the other one uses only one global network. To avoid any confusion, we refer to the algorithm with multiple networks as Method~I, and the global one as Method~II.

\subsection{Method~I: multiple neural networks}

\subsubsection{The regression rule}
By taking the expectation of the discounted option value conditioned on $\mathcal{F}_{t_i}$ and applying the Doob-Meyer decomposition, we have
\begin{equation}\label{eq: RegressionBasis}
V_{t_{i+1}}=\EV{V_{t_{i+1}}\Big|\FF_{t_{i}}}+\int_{t_{i}}^{t_{i+1}} H_{u} \D W_{u}.
\end{equation}
In this equation, the conditional expectation is the \emph{continuation value}, and the integral is the \emph{martingale increment} from $t_{i}$ to $t_{i+1}$. Since the stock price process is Markovian, both the conditional expectation and the process $(H_{t_i})_{i \in \{0, 1, ..., n\}}$ can be estimated as functions of the state variables $S_{t_i}$ \citep{ccinlar1980semimartingales, chitashvili1997functions}.

Let $\Phi^{i}(S_{t_i}): \mathbb{R}^d\to\mathbb{R}$ and $\Psi^{i}(S_{t_i}): \mathbb{R}^d\to\mathbb{R}^d$ be approximations of the continuation function and the process $H_{t_i}$ at $t_i$, respectively. We refer to $\Psi(S_{t_i})$ as the \textit{martingale increment function}. Due to the independence among stocks, the martingale increment can be approximated by 
$\sum_{j=1}^{d}\beta_{t_{i}}^{\-1}\Psi^{i}(S_{t_{i}})^{j}\Delta W_{t_{i}}^{j}$, where $\Delta W_{{t_i}}^{j} = W_{{t_{i+1}}}^{j} - W_{{t_i}}^{j}$. For simplicity, we leave out the superscript that indicates dimension and the symbol for summation in the rest of this section.

Based on \eqref{eq: RegressionBasis}, we perform regression at each time through: 
\begin{equation*}
    \min_{\Phi,\Psi} \left(\beta_{t_{i}}V_{t_{i+1}}-\Phi^{i}(S_{t_i})-\sum_{j=1}^{d} \Psi^{i}(S_{t_{i}})^{j}\Delta W_{t_{i}}^{j}\right)^{2}.
\end{equation*}

In this method, one neural network is used to regress the continuation value and the martingale increment on the current stock prices at time $t_i \in \{t_0, t_1, t_2, ..., t_{n-1}\}$. 
In this work, we use fully-connected feedforward neural networks to perform these regressions. Let $N\hspace{-0.12em}N^\Theta$ denote fully connected feedforward artificial neural networks, with $\Theta$  describing the structure of a network. $\Theta=(L, [n_{1}, \ldots, n_{L}])$ represents a network with $L$ layers, and each layer $l$ has $n_{l}$ neurons. In particular, $n_{1}$ and $n_{L}$ are the number of input features and the number of outputs, respectively. Each network takes the form: 
$$\sigma_{n_{L-1}}\circ A_{L-1}\circ\cdots\circ\sigma_{1}\circ A_1,$$
where $A_{l}$ signifies an affine transformation from layer $l$ to layer $l+1$:
$$A_l(x)=w_l^T x+b_{l},$$for $x\in\RR^{n_l}$, $w_l\in \RR^{n_l\times n_{l+1}}$, $b_l\in\RR^{l+1}$, and $\sigma_{l}$ is the activation function applied to $A_{l}$.

\subsubsection{The stopping strategy}
Let $\tau_{i}: \Omega\to\{t_{i+1}, \ldots, t_{n}=T\}$ be a stopping time, and $\Tau_{i}$ be the set of all stopping times that takes values greater than $t_{i}$.  

The optimal stopping strategy is to exercise the option once the immediate payoff is higher than the continuation value. Let $f(S_{t}):\mathbb{R}^d \to \mathbb{R}$ be the payoff of the option at $t\in[0, T]$.
The stopping time can be represented as: 
$$\tau_{i} = \min\{t_j\in\{t_{i+1}..., t_{n-1}\}: f(S_{t_j}) \geq \Phi^{j}(S_{t_j})\}\wedge t_{n}.$$ 

\subsubsection{The update rule}
Consider two random processes $(Y_{t_i})^n_{i=1}$ and $(X_{t_i})^n_{i=1}$, defined as the following: 
\begin{itemize}
    \item At $t_n=T$, the option holder has to either exercise the option if it is in the money or let it expire if it is out of the money. Let 
     $$Y_{t_n}=X_{t_n}=f(S_{t_n}).$$
    \item At each $t \in \{t_{0}, t_{1}, \ldots, t_{n-1}\}$, the option holder either exercises the option immediately if the payoff value is higher than the continuation value, or hold it till the next exercise point if it is lower. Let
\begin{equation*}
 Y_{t_i}=\begin{cases}
     f(S_{t_i}) &\text{, if } f(S_{t_i})\geq\Phi^{i}(S_{t_i})\\
     \beta_{\Delta t}^{-1} Y_{t_{i+1}}-\Psi^{i}(S_{t_i}) \cdot\Delta W_{t_i} &\text{, if } f(S_{t_i})<\Phi^{i}(S_{t_i}).
 \end{cases}
\end{equation*}
\begin{equation}\label{eq: X_updates}
    X_{t_i}=\begin{cases}
        f(S_{t_i})  &\text{if } 
        f(S_{t_i}) \geq \beta_{\Delta t}^{-1} X_{t_{i+1}}-\Psi^{i}(S_{t_i})\cdot\Delta W_{t_i}\\
        \beta_{\Delta t}^{-1}X_{t_{i+1}}-\Psi^{i}(S_{t_i})\cdot\Delta W_{t_i}, &\text{if } 
        f(S_{t_i}) < \beta_{\Delta t}^{-1} X_{t_{i+1}}-\Psi^{i}(S_{t_i})\cdot\Delta W_{t_i}
    \end{cases}.
\end{equation}
\end{itemize}

In the update of $Y_{t_i}$, neglecting the subtraction term, it simply applies the stopping strategy. By averaging $Y_{t_{0}}$ over all paths, we get a lower bound of the option price. The discounted payoff at the optimal stopping time is used as the regression target, which can significantly reduce the bias but introduce a higher variance. To cancel this negative effect, we subtract the martingale increment $\Psi^{i}(S_{t_i})\cdot\Delta W_{t_i}$ adjusted with the time value. If the approximation of $H_{t_i}$ is perfect, the variance can be cancelled out completely. A proof is given in Appendix \ref{app: VR} to show that this term reduces the variance of the estimate. 

In the update of $X_{t_i}$, the subtraction of the martingale increment serves a different purpose. Note that \eqref{eq: X_updates} can be written as  a recursive equation:
$$X_{t_{i}} = \max \{ f(S_{t_i}), \; \beta_{\Delta t}^{-1} X_{t_{i+1}} \!-\! \Psi^{i}(S_{t_i})\cdot\Delta W_{t_i} \}.$$
By expanding the recursion, we can have: 
\begin{align}\label{eq: X0_recursive}
    X_{t_{0}} 
    =  \max \{ f(S_{t_0}), \; 
        \beta_{\Delta t}^{-1} \max \{ f(S_{t_{1}}), \; \ldots, & \\  
         \beta_{\Delta t}^{-1}  \max \{ f(S_{t_{n\-1}}), 
          \beta_{\Delta t}^{-1} f&(S_{t_{n}})  
            \!-\! \Psi^{n\-1}(S_{t_{n\-1}}) \!\cdot\! \Delta W_{t_{n\-1}} \} \nonumber \\ 
              \cdots 
        \!-\!  \Psi^{1}& (S_{t_{1}}) \!\cdot\! \Delta W_{t_{1}} \} 
       \!-\! \Psi^{0}(S_{t_{0}}) \!\cdot\! \Delta W_{t_{0}} \}. \nonumber
\end{align}
Recall the duality \eqref{eq: Duality}. For a martingale $M$, we have: 
\begin{align} \label{eq: DeriveX}
V_{t_{0}} 
\leq & \max\{Z_{t_0}, \ldots, Z_{t_{n-1}} \!-\! \sum_{i=0}^{n-2} \Delta M_{t_{i}}, Z_{t_{n}} \!-\! \sum_{i=0}^{n-1} \Delta M_{t_{i}}\}  \nonumber \\
=& \max\{Z_{t_0}, \ldots,  \max\{Z_{t_{n-1}} ,Z_{t_{n}} \!-\! \Delta M_{t_{n-1}}\} \!-\! \sum_{i=0}^{n-2} \Delta M_{t_{i}}\} \nonumber \\
=&\max\{Z_{t_0}, \max\{Z_{t_{1}}, \ldots
   \max\{Z_{t_{n-2}}, \max\{Z_{t_{n-1}}, Z_{t_{n}}\!-\! \Delta M_{t_{n-1}} \}- \Delta M_{t_{n-2}}\}\nonumber \\
 & \qquad \qquad \qquad \qquad \;
   \ldots \!-\! \Delta M_{t_{1}}\} \!-\! \Delta M_{t_{0}}\}.
\end{align} 
Since 
$\Delta M_{t_{i}} \approx \beta_{t_{i}}^{\-1}\Psi^{i}(S_{t_{i}})\Delta W_{t_{i}}$ and $Z_{t_{i}}=\beta_{t_{i}}^{-1}f(S_{t_{i}})$, from \eqref{eq: X0_recursive} and \eqref{eq: DeriveX}, we can see that $\EV{X_{t_{0}}}$ is an upper bound of the price. 

The processes $X_{t_i}$ and $Y_{t_i}$ can also be interpreted in the following way. The variable $Y_{t_i}$ is a proxy of the buyer's price, as the two cases correspond to the stopping decision based on comparing the exercise payoff and the continuation value. The variable $X_{t_i}$ is a proxy of the seller's price, as the two cases correspond to whether the seller needs to update their hedging targets based on the comparison of the exercise payoff and the hedging price.

Note that in all numerical experiments shown in this paper, $t_{0}$ is not considered an exercise date, coinciding with the fact that one does not exercise the option at the initial time. This choice is reflected in the algorithm \ref{alg: MultiNN} by directly letting  $Y_{t_{0}} = \beta_{\Delta t}^{-1} Y_{t_{1}}-\Psi^{0}(S_{t_{0}}) \cdot \Delta W_{t_{0}}$ without checking the comparison condition. However, the regression is still performed to obtain the martingale increment function at $t_{0}$.

\subsubsection{The whole process}
We outline the entire pricing process using Method~I in the algorithm below. Throughout the training process, all trained models are saved for future use. Subsequently, an independent out-of-sample simulation is conducted to derive estimates. This second simulation can be executed in two ways: following the training algorithm by determining the values backward, or starting from the initial time and making decisions forward. 

\begin{algorithm}[H]
\DontPrintSemicolon
\SetAlgoLined
\KwResult{Functions $\Phi^{i}$, $\Psi^{i}$ for $i\in\{0, 1, ..., n-1\}$ }
Simulate $N$ stock paths\;
Initialise $Y_{t_n}=X_{t_n}=\max(f(S_{t_n}),0)$\; 
 \For {i=n-1:1}{
  Regress $\beta_{\Delta t}^{-1} Y_{t_{i+1}}$ on $S_{t_i}$: $\underset{\Phi^{i},\Psi^{i}}{\min}(\beta_{\Delta t}^{-1}Y_{t_{i+1}} -\Phi^{i}(S_{t_i})-\Psi^{i}(S_{t_i}) \Delta W_{t_i})^2$\;
  $Y_{t_i}=\beta_{\Delta t}^{-1}Y_{t_{i+1}}-\Psi^{i}(S_{t_i}) \Delta W_{t_i}$\;
  $X_{t_i}=\beta_{\Delta t}^{-1}X_{t_{i+1}}-\Psi^{i}(S_{t_i})\Delta W_{t_i}$\;
  \If{$ f(S_{t_i}) > \Phi^{i}(S_{t_i})$}{
   $Y_{t_i}=f(S_{t_i})$\;
   }
  \If{$f(S_{t_i}) > X_{t_i}$}{
   $X_{t_i}=f(S_{t_i})$\;
   }
 }
 Regress $\beta_{\Delta t}^{-1}Y_{t_1}$ on $S_{t_0}$: $\min(\beta_{\Delta t}^{-1}Y_{t_1} -\Phi^{0}(S_{t_0})-\Psi^{0}(S_{t_0}) \Delta W_{t_0})^2$\;
  $Y_{t_0}=\beta_{\Delta t}^{-1}Y_{t_{1}}-\Psi^{0}(S_{t_0}) \Delta W_{t_0}$\;
 $X_{t_0} = (\beta_{\Delta t}^{-1}X_{t_{1}}-\Psi^{0}(S_{t_0}) \Delta W_{t_0})
        \indic_{f(S_{t_{0}}) \leq X_{t_{0}}}
        +f(S_{t_{0}})\indic_{f(S_{t_{0}}) > X_{t_{0}}}$\;
 \caption{American-style Option Pricing with Multiple Neural Networks\label{alg: MultiNN}}
\end{algorithm}

\subsubsection{Discussion on the convergence}
Since its introduction, numerous studies have been conducted to analyse the convergence analysis of the LSMC method. In their original work, \citet{longstaff2001} showed the convergence in cases with only two early exercise opportunities. Subsequently,  \citet{clement2002analysis} established a more general almost sure convergence by modifying the method to regress all paths instead of solely in-the-money ones.
\citet{egloff2005monte} showed both the convergence and error estimates by using Vapnik-Chervonenkis classes with the assumption of convexity, allowing for relaxation of linearity assumption in approximation spaces.
Eventually, \citet{zanger2018convergence} derived a general convergence result, providing new overall error estimates for the algorithm without assuming linearity or convexity of approximation spaces, and without requiring an independent data set. This result validated the application of neural networks in the method and the avoidance of an independent data set in regression. 
Regarding the upper bound, \citet{hure2021convergence} established the convergence of the method for deriving upper bounds in the context of solving reflected BSDEs. 

The convergence of our method, which incorporates duality into the primal problem, follows from the literature mentioned above.  

\begin{remark}
\citet{el1997american} showed that pricing American options is related to reflected BSDEs, the solution of which is an $\mathcal{F}_t$-measurable triple $(V_{t}, H_{t}, K_{t})$ for $t \in [0,T]$ with values in $(\RR, \RR^{n}, \RR_{+})$, and satisfies:
\begin{equation*}
\begin{cases}
V_{t} = Z_{T} + \int_{t}^{T} b(s, V_{s}, H_{s})\D s + K_{T} - K_{t} -\int_{t}^{T} H_{s} \D W_{s},\\
V_{t} \geq Z_{t},\quad 0\leq t\leq T,\\
K_{0} = 0,\text{ and }\int_{0}^{T} (V_{t}-Z_{t})\D K_{t} = 0.
\end{cases}.
\end{equation*}
Our work can be easily extended to solve this type of BSDE. The processes $V$ and $H$ here have the same meaning as we have defined before, and our work generates numerical solutions for them. The process $K$ can be seen as the non-decreasing process $A$ and calculated by a second simulation where we accumulate the gap between the value process and the payoff process. Note we have $b(\cdot, \cdot, \cdot)=0$ in our case. However, if we have a model where $b(\cdot, \cdot, \cdot)\neq0$, we can still approximate it by adding one more term to our regression.
\end{remark}

\subsection{Method~II: one global neural network}
\begin{figure}[ht]
    \centering
    \includegraphics[width=0.72\paperwidth]{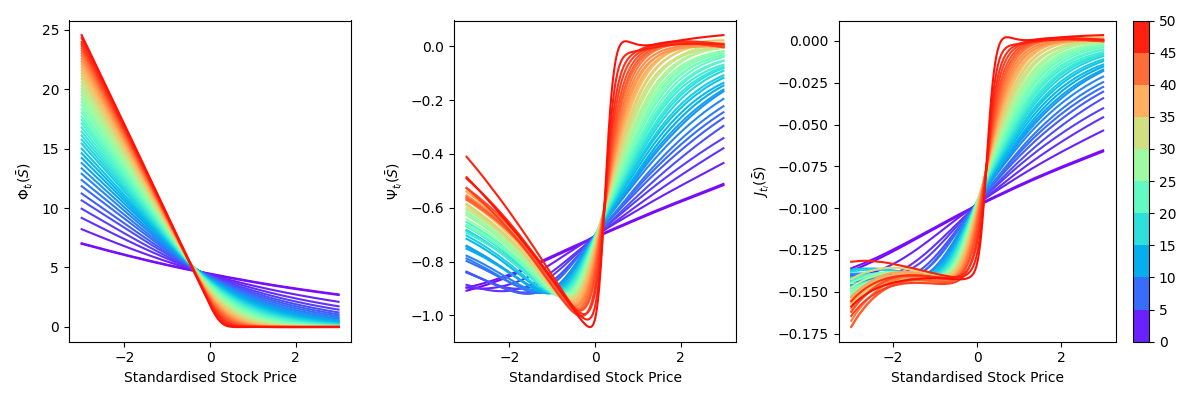}
    \caption{Estimates of continuation functions, martingale increment functions and hedging ratio of a 1D American-style put option with $50$ exercise dates (same parameters as the one in Section \ref{sec: numerical_results}). Each line represents a function at a step. \textit{Left}: the continuation function; \textit{Middle}: the martingale increment function; \textit{Right}: the hedging ratio. The colorbar represents the step: 0 is the initial time and 50 is the maturity.}
    \label{fig: FuncsPlot}
\end{figure}

After pricing a vanilla American-style put option under the Black-Scholes model that has $50$ exercise points using Method~I, we plot $\Phi^{i}(\Bar{S}_{t_i})$, $\Psi^{i}(\Bar{S}_{t_i})$, and the hedging ratio $J^{i}(\Bar{S}_{t_i})$ for $i\in \{0, 1, ..., 49\}$, in Figure \ref{fig: FuncsPlot}, to visualise the
approximated functions, where $\Bar{S}_{t_i}$ is the standardised stock price. We can see that continuation functions and the martingale increment functions at different times $t_i$ have similar shapes, and they evolve continuously in time.

\begin{remark}
Under the Black-Scholes model, the first derivative of a continuous function is expected to align with the hedging ratio, thereby establishing a link to the martingale increment function. While we might expect the martingale increment functions to present a flat trajectory near the value of $-1$ along the left axis, the middle plot in Figure \ref{fig: FuncsPlot} displays a deviation from this pattern. This discrepancy arises because the plot illustrates the approximated martingale increment function of the standardised stock price, rather than the direct hedging ratios with respect to the stock price itself. By adjusting $\Psi^{i}$ with the diffusion term, we can illustrate the hedging ratio, shown on the right plot, and it is more aligned to the expected shape. Furthermore, we have not imposed any constraints restriction on the shape of the function during training. The further the data points deviate from the centre, the less data is available, leading to increased extrapolation at the plot's extremities.  
\end{remark}

\subsubsection{The whole process}
Based on the similarity in the shape of functions and their continuous progression in time, we propose a second method where we only use one network for all regressions by including the time/step as an input variable. 

We apply the same stopping strategy, and the regression and the updates of process $X$ and $Y$ at each time remain the same. However, this approach poses additional challenges as it requires target values at all times when we start training the model. In method~I, the update of $Y_{t_i}$ before the regression provides a relatively accurate target values for the training of the corresponding network, but this is not available in method~II. To overcome this challenge, we propose a novel approach where we achieve the goal by alternating the model training and stopping strategy updates. 

Initially, we set the maturity as the stopping time, so target values at $t_i \in \{t_{0}, \ldots, t_{n-1}\}$ are $\beta_{(n-i)\Delta t}^{-1}f(t_{n})$. We train the model using these target values for a given number of epochs and then use the trained model to determine a new series of $Y_{t_{i}}$ using the update rule stated before. Once all target values are updated, we do the training again. We repeat this training-updating process till some predefined criterion is met. We choose small numbers as the number of epochs among updates, especially for the earlier training, since the stopping strategies we applied are unlikely to be optimal at the start.  

Denote $\Phi_{\text{II}}(t_i, S_{t_i}): \RR_+\times\RR^d\longrightarrow\RR$ and $\Psi_{\text{II}}(t_i, S_{t_i}): \RR_+\times\RR^d \longrightarrow\RR^d$ as the approximations of the continuation functions and the martingale increment functions. Method~II is summarised in Algorithm~\ref{alg: OneNN}.

\begin{algorithm}[H]
\DontPrintSemicolon
\SetAlgoLined
\KwResult{Functions $\Phi_{\text{II}}$, $\Psi_{\text{II}}$ }
 Simulate $N$ stock price paths\;
 Initial: $Y_{t_{i+1}}=\beta_{(n-i+1)\Delta t}^{-1} f(S_{t_n})$, for $i \in \{0, ..., n-1\}$\;
 \While {$\mathrm{stopping\ criterion\ is\ not\ met}$}{
 \For {$i=1:\mathrm{epoch}$}{
  
  Regress $\beta_{\Delta t}^{-1}Y_{t_{i+1}}$ on $(t_i, S_{t_i}, \Delta W_{t_i})$ for $i \in \{0, ..., n-1\}$: 
  
  $\underset{\Phi_{\text{II}},\Psi_{\text{II}}}{\min}\!\left(\beta_{\Delta t}^{-1}Y_{t_{i+1}}-\Phi_{\text{II}}(t_i, S_{t_i})-\Psi_{\text{II}}(t_i, S_{t_i}) \Delta W_{t_i}\right)^2$  \
  
 }
 $Y_{t_n} = X_{t_n} =f(S_{t_n})$\;
 \For {$i=n-1:1$}{
   $Y_{t_i}=\beta_{\Delta t}^{-1}Y_{t_{i+1}}-\Psi(t_i, S_{t_i})\Delta W_{t_i}$\;
   \If{$f(S_{t_i})>\Phi_{\text{II}}(t_i, S_{t_i})$}{
   $Y_{t_i} = f(S_{t_i})$\;
   }
   $X_{t_i} = \beta_{\Delta t}^{-1}X_{t_{i+1}}-\Psi_{\text{II}}(t_i, S_{t_i})\Delta W_{t_i}$\;
   \If{$f(S_{t_i})>X_{t_i}$}{
   $X_{t_i}=f(S_{t_i})$
   }
  }
  $Y_{t_0}=\beta_{\Delta t}^{-1}Y_{t_{1}}-\Psi_{\text{II}}(t_0, S_{t_0})\Delta W_{t_0}$\;
  $X_{t_0}=(\beta_{\Delta t}^{-1}X_{t_{1}}-\Psi_{\text{II}}(t_0, S_{t_0})\Delta W_{t_0})
            \indic_{f(S_{t_0}) \leq X_{t_0}}
            +f(S_{t_0})\indic_{f(S_{t_0}) > X_{t_0}}$\;
}
 \caption{American-style Option Pricing with One Global Network
 \label{alg: OneNN}}
\end{algorithm}

\subsubsection{Discussion on the convergence}
While single global networks have been used to address a wide range of optimal stopping problems, the convergence analysis for backward methods remains lacking. \citet{vanroy2001, herrera2021optimal} have proposed similar approaches to price American-style options with some insights into the convergence properties. A key distinction between our method to theirs is that they use estimated continuation value rather than the exact optimal payoff to make stopping decisions, and they start with a completely random initial strategy. They showed that the method converges eventually, but there are no results on the rate of convergence and the error bounds. 

The rationale behind our proposed method is that when the initial strategy is to wait until maturity, we have relatively accurate target values for decisions made closer to the maturity date. In particular, the target values for determining the second-to-last exercise decisions would be exact. The use of the exact optimal payoff avoids the reliance on estimated continuation values for making stopping decisions. As training progresses, the stopping decisions at later times will improve first, which in turn, improves the accuracy of target values for earlier stopping decisions. This iterative refinement ensures that, over time, the target values for all times converge to their true values, enhancing the overall decision-making process.

\section{Algorithm variants}
There are two sources of errors in our methods. Firstly, there is the time discretisation error induced by approximating the continuous martingale using the Euler scheme. This error scales proportionally with the step size square root $\sqrt{\Delta t}$, potentially resulting in suboptimal upper bounds in cases where the option offers infrequent exercise opportunities. The other source is regression, which can be mitigated by using a larger data set, utilizing more suitable network architectures, and prolonging the training duration.
However, these approaches come at the expense of increased computational costs and memory requirements.

To enhance the performance of our algorithms, we present five different variants aimed at generating more accurate results, reducing computational cost and addressing memory exhaustion issues.  In this section, we present numerical results to evaluate the effectiveness of each variant in pricing 1D put options, 5D max-call options, or both. These options share parameters with those presented in Section \ref{sec: numerical_results}. The objective of this section is to assess the impact of different variants on our methods through comparisons with the original version.

\subsubsection*{Variation 1: add a second term for martingale increment approximation}
The approximation 
$\sum_{j=1}^{d} \Psi^{i}(S_{t_{i}})^{j}\Delta W_{t_{i}}^{j}$ 
deteriorates with the step size increasing. To improve the accuracy of the martingale increment estimates, we propose to add one more term in the regression.
The choice of the term depends on the model, provided it satisfies the martingale property of having a zero mean increment. We choose $(\Delta W_{t_{i}})^2-\Delta t$ in our work, which can be connected to the Milstein scheme. This variant requires more outputs from the network and results in a change of the loss function: 
\begin{equation*}
    \min_{\Phi,\Psi_1, \Psi_2} \left(\beta_{\Delta_t}^{-1}Y_{t_{i+1}}
    -\Phi(S_{t_i})
    -\sum_{j=1}^{d}\Psi_{1}^{i}(S_{t_i})^{j}\Delta W_{t_i}^{j} 
    -\sum_{j=1}^{d}\Psi_{2}^{i}(S_{t_i})^{j}((\Delta W_{t_i}^{j})^2-\Delta t)\right)^2.
\end{equation*}
The updates of $X$ and $Y$ need to be changed accordingly as the martingale increment becomes the sum of two terms. This variation can be applied to both methods.

The changes in results introduced by this variation applied to method~I are shown in Table~\ref{tab: var_2ndMG}. We can see that with similar training times in each case, this variant significantly reduces the gap between the lower and the upper bound, mainly caused by better approximations of the upper bound. The lower bound also improves due to a more effective variance reduction. Additionally, this improvement is more pronounced when the pricing problem is more complicated.
\begin{table}[ht]
  \centering
\begin{tabular}{|c|c|c|c|c|c|c|c|c|}
    \hline
          \multicolumn{3}{|l}{} & \multicolumn{2}{|c}{LB} & \multicolumn{2}{|c}{UB} & \multicolumn{2}{|c|}{Diff} \\
    \hline
    \multicolumn{2}{|l|}{} & \multicolumn{1}{l|}{Time} & \multicolumn{1}{l|}{Mean} & \multicolumn{1}{l|}{S.D.} & \multicolumn{1}{l|}{Mean} & \multicolumn{1}{l|}{S.D.} & \multicolumn{1}{l|}{Mean} & \multicolumn{1}{l|}{S.D.} \\ 
    \hline
    \multirow{2}[0]{*}{1D} & 1 Term  & 34    & 4.4748 & 0.0007 & 4.5559 & 0.0022 & 0.0811 & 0.0024 \\
    \cline{2-9}
          & 2 Terms   & 34    & 4.4765 & 0.0002 & 4.4936 & 0.0017 & 0.0171 & 0.0018 \\
    \hline
    \multirow{2}[0]{*}{5D} & 1 Term  & 56    & 26.1372 & 0.0090 & 28.2132 & 0.0169 & 2.0761 & 0.0177 \\
    \cline{2-9}
          & 2 Terms  & 52    & 26.1464 & 0.0039 & 26.8974 & 0.0074 & 0.7510 & 0.0081 \\
    \hline
    \end{tabular}%
  \caption{Option Pricing with/without a second martingale increment term applied in method~I. The first column indicates the option type, and the second column shows whether a second term is added in the martingale increment approximation. The networks in the 1D case have three hidden layers of 30 neurons, and the one in the 5D case has two layers of 64 neurons.}
  \label{tab: var_2ndMG}%
\end{table}%

\subsubsection*{Variation 2: add sub-steps}
\begin{figure}[ht]
\centering
\begin{tikzpicture}[yscale=0.9, xscale=0.8]
\begin{axis}[
    title={Price Bounds},
    xlabel={Number of Substeps},
    ylabel={Estimate},xmin=0, xmax=32, ymin=26, ymax=27,
    ytick={26, 26.5, 27},
    legend pos=north east
]
\addplot table [x=sub_step2, y=lb_B_mul, col sep=comma] {DataForPlots.csv};
\addplot table [x=sub_step2, y=ub_B_mul, col sep=comma] {DataForPlots.csv};
\addplot table [x=sub_step3, y=lb_B_one, col sep=comma] {DataForPlots.csv};
\addplot table [x=sub_step3, y=ub_B_one, col sep=comma] {DataForPlots.csv};
\legend{LB-Method~I, UB-Method~I, LB-Method~II, UB-Method~II}
    
\end{axis}
\end{tikzpicture}
\begin{tikzpicture}[yscale=0.9, xscale=0.8]
\begin{axis}[
    title={Running Time},
    xlabel={Number of Substeps},
    ylabel={Estimate},xmin=0, xmax=32, ymin=0, ymax=3000,
    ytick={0, 1000, 2000, 3000},
    legend pos=north east
]
\addplot table [x=sub_step2, y=time_B_mul, col sep=comma] {DataForPlots.csv};
\addplot table [x=sub_step2, y=time_B_one, col sep=comma] {DataForPlots.csv};
\legend{Method~I, Method~II}
    
\end{axis}
\end{tikzpicture}
\caption{Price bounds (\textit{Left}) and corresponding running times (\textit{Right}) of a 5D max-call Bermudan option with different numbers of substeps using both method~I and II.}
\label{fig: var_substep}
\end{figure}
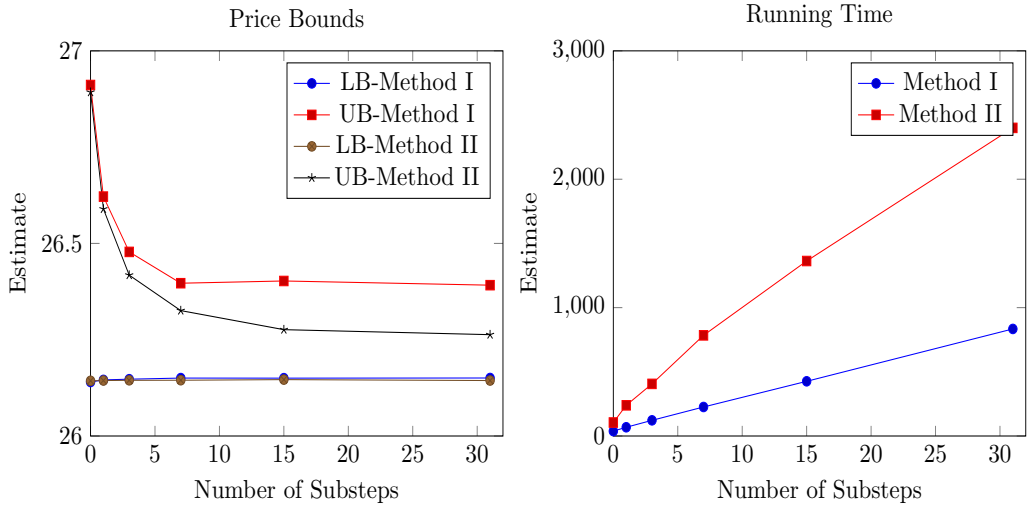

Refining the martingale approximation can also be achieved by reducing the step size. In the context of Bermudan option pricing, this can be achieved by adding substeps between two exercise times, where we do not make stopping decisions but only accumulate martingale increments. This variant is particularly important in pricing options with less frequent exercise opportunities. The 5D max-call option we have been pricing has only $9$ exercise opportunities over $3$ years. As demonstrated in Figure \ref{fig: var_substep}, adding substeps markedly enhances the accuracy of the upper bound estimation when pricing this option.  The initial introduction of substeps brings about a notably sharper improvement in the bounds, which tends to taper off as more substeps are added. 
However, it is important to note that the computational time increases with the addition of substeps, as also depicted in Figure \ref{fig: var_substep}, but the speed of increase is slower than linear.

Figure \ref{fig: var_substep} also indicates that Method~II produces better results with slower training, but this observation can vary with the adjustments of training parameters. Further comparisons between these two methods will be detailed throughout this paper, forming a conclusion at the end.

\subsubsection*{Variation 3: use separate networks for the two functions}
In our approaches, we initially utilized a single network to estimate both the continuation value function and the martingale increment function. However, given the potential complexity difference between these functions, especially when the model gets more complicated and the dimension gets higher, we propose to use separate networks to approximate them, where one is dedicated to generating the continuation value and the other for calculating the martingale increment functions. To evaluate the efficacy of this variant, we applied it to Method~I across three different scenarios: a 1D put, a 5D max-call with no substep, and a 5D max-call with 31 substeps. For each scenario, we ensured that the networks had a comparable number of parameters. We can see from Table \ref{tab: var_seperate_nn} that variant 3 can produce more accurate results with less training time in all three cases, and this effect is more notable in more complex problems (5D max-call option with 32 substeps). When implementing this variant with Method~II, we observed a similar pattern, reinforcing the benefits of employing separate networks for approximating distinct functions.

\begin{table}[htbp]
  \centering
    \begin{tabular}{|c|c|c|c|c|c|c|c|c|}
    \hline
          \multicolumn{3}{|l}{} & \multicolumn{2}{|c}{LB} & \multicolumn{2}{|c}{UB} & \multicolumn{2}{|c|}{Diff} \\
    \hline
    \multicolumn{1}{|l|}{} & \multicolumn{1}{l|}{Separate} & \multicolumn{1}{c|}{Time} & \multicolumn{1}{c|}{Mean} & \multicolumn{1}{c|}{S.D.} & \multicolumn{1}{c|}{Mean} & \multicolumn{1}{c|}{S.D.} & \multicolumn{1}{c|}{Mean} & 
    \multicolumn{1}{c|}{S.D.} \\    
    \hline
    \multirow{2}[0]{*}{1D} & False & 33    & 4.4766 & 0.0001 & 4.4929 & 0.0014 & 0.0162 & 0.0014 \\
    \cline{2-9}
     & True& 29    & 4.4757 & 0.0005 & 4.4899 & 0.0009 & 0.0141 & 0.0009 \\
    \hline
    \multirow{2}[0]{*}{5D0S} & False & 65    & 26.1318 & 0.0054 & 26.9200 & 0.0070 & 0.7883 & 0.0105 \\
    \cline{2-9}
    & True& 64    & 26.1138 & 0.0066 & 26.8869 & 0.0055 & 0.7731 & 0.0073 \\
    \hline
    \multirow{2}[0]{*}{5D31S} & False &1179  & 26.1528 & 0.0012 & 26.2887 & 0.0110 & 0.1359 & 0.0113 \\
    \cline{2-9}
    & True& 994   & 26.1527 & 0.0011 & 26.2263 & 0.0028 & 0.0736 & 0.0026 \\
    \hline
    \end{tabular}%
  \caption{Options pricing with/without using separate networks in Method~I. In the first column, 1D, 5D0S, and 5D31S represent the option priced: the 1D American-style put, the 5D max-call Bermudan with no substep and the 5D max-call Bermudan with $31$ substeps. The second column indicates whether separate networks are used.}
  \label{tab: var_seperate_nn}%
\end{table}%

\subsubsection*{Variation 4: train on data from parts of the exercise times}
In Method~II, the standard practice involves training the model across all simulated paths at every timestep. Anticipating that data shares similarities across different times, we suggest an alternative strategy that focuses on training with data from selectively chosen timesteps. This approach hinges on the premise that not every timestep contributes uniquely to model accuracy, allowing for strategic data reduction. Two methodologies are proposed for selecting which timesteps to include in the training process: a random selection or a systematic, evenly-spaced grid approach. For example in a scenario with 50 exercise opportunities and the aim is to train the model using data from only half of the exercise times, we could either randomly choose $25$ timesteps from the set $\{0, 1, 2, ..., 49\}$ or use data from every other timestep, i.e. $t=t_{1}, t_{3},..., t_{49}$.

\begin{figure}[ht]
\centering
\begin{tikzpicture}[yscale=0.9, xscale=0.8]
\begin{axis}[
    xlabel={Number of steps used},
    ylabel={Training Time (Secs)},
    xmin=0, xmax=50,
    ymin=200, ymax=700,
    ytick={200, 400, 600, 700},
    legend pos=north west
]

\addplot table [x=t_trained, y=time_grid, col sep=comma] {DataForPlots.csv};
\addplot table [x=t_trained, y=time_rand, col sep=comma] {DataForPlots.csv};
\legend{Grid, Random}  

\end{axis}
\end{tikzpicture}
\begin{tikzpicture}[yscale=0.9, xscale=0.8]
\begin{axis}[
    xlabel={Number of steps used},
    ylabel={Bounds Difference},xmin=0, xmax=50,ymin=0.013, ymax=0.028,
    ytick={0.013, 0.018, 0.023, 0.28},
    legend pos=north east
]

\addplot table [x=t_trained, y=diff_grid, col sep=comma] {DataForPlots.csv};
\addplot table [x=t_trained, y=diff_rand, col sep=comma] {DataForPlots.csv};
\legend{Grid, Random} 

\end{axis}
\end{tikzpicture}
\caption{Changes in the estimated bounds and training time with different numbers of timesteps used in training. The option priced is a 1D American-style put option.}
\label{fig: var_part_time}
\end{figure}
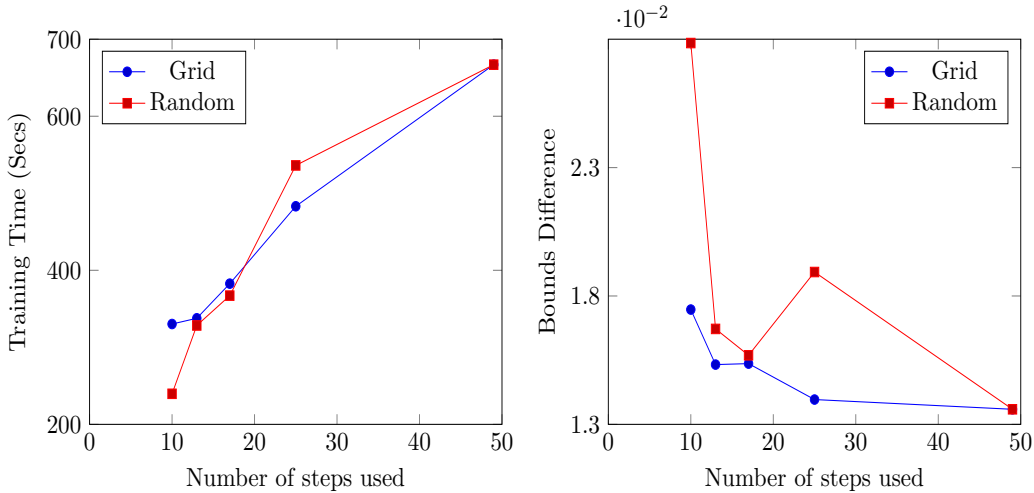

In Figure \ref{fig: var_part_time}, we illustrate the impact of this timestep selection strategy on training duration and the accuracy of the results when pricing a 1D American-style put. This modification clearly reduces the computational cost but also compromises the accuracy of the results. While this trade-off is anticipated, our goal is to strike a balance between computational efficiency and result accuracy.

\subsubsection*{Variation 5: generate fresh data while training}
In general, larger training sets often yield more accurate and robust results, albeit at the expense of increased computational demands. However, due to the nature of the problem, we have to simulate the whole path before the training. The memory requirement can become extremely high, particularly in high-dimensional problems. To address this challenge, \citet{chan2006pricing, aid2014probabilistic} recommended storing the random seed used during simulation. This enables us to only preserve data points at one step in a path and discard the remainder. Once the network is trained, the state values for subsequent networks are reconstructed using the saved seed and the current states. However, this introduces more calculation during training and can only be applied to Method~I. To overcome these limitations, we introduce an alternative solution designed to circumvent the memory constraints in scenarios where Method~II is employed, where data from all times are needed when training starts.

In the original Method~II, all $N$ paths are generated at the start. Among updates, the data set is split into the training and the validation set randomly and the training set is then grouped into batches of size $N_{batch}$. The network is then trained for a given number of epochs by looping over all training batches in each epoch. The validation set is then used to check the stopping criteria after each update. With this variant, we only generate the validation set before the start of the training, serving the same purpose as in the original version. Among updates, we generate $N_{batch}$ paths, and train the network using this batch for a given number of epochs, and then discard them. We repeat this generating-training-discarding process multiple times before the stopping criteria are evaluated. By utilizing smaller batch sizes and continually generating additional paths as needed, we can effectively train on a larger number of paths without encountering memory exhaustion issues. 

\begin{figure}[ht]
\centering

\begin{tikzpicture}[yscale=0.9, xscale=0.8]
\begin{axis}[cycle list name=exotic,
    xlabel={Time(secs)},
    ylabel={Bounds Differences},
    xmin=0, xmax=520,
    ymin=0, ymax=1,
    ytick={0, 0.5, 1},
    legend pos=north east
]

\addplot table [color=blue, x=time_1_1d, y=diff_1_1d, col sep=comma] {DataForPlots.csv};
\addplot table [color=red, x=time_5_1d, y=diff_5_1d, col sep=comma] {DataForPlots.csv};
\addplot table [color=green, x=time_10_1d, y=diff_10_1d, col sep=comma] {DataForPlots.csv};
\addplot table [color=black, x=time_25_1d, y=diff_25_1d, col sep=comma] {DataForPlots.csv};
\addplot table [color=brown, x=time_50_1d, y=diff_50_1d, col sep=comma] {DataForPlots.csv};
    \legend{1 batch, 5 batches, 10 batches, 25 batches, 50 batches}
\end{axis}
\end{tikzpicture}
\begin{tikzpicture}[yscale=0.9, xscale=0.8]
\begin{axis}[
    xlabel={Time(secs)},
    ylabel={Bounds Differences},
    xmin=0, xmax=520,
    ymin=0, ymax=1,
    ytick={0, 0.5, 1},
    legend pos=north east
]

\addplot table [x=time_r, y=diff_r, col sep=comma] {DataForPlots.csv};
\addplot table [x=time_25_1d, y=diff_25_1d, col sep=comma] {DataForPlots.csv};
\addplot table [x=time_multi, y=diff_multi, col sep=comma] {DataForPlots.csv};
\legend{Method~II, Method~II-V5, Method~I}
\end{axis}
\end{tikzpicture}

\begin{tikzpicture}[yscale=0.9, xscale=0.8]
\begin{axis}[cycle list name=exotic,
    xlabel={Time(secs)},
    ylabel={Bounds Differences},
    xmin=0, xmax=1020,
    ymin=0, ymax=4,
    ytick={0, 1, 2, 3, 4},
    legend pos=north east
]

\addplot table [color=blue, x=time_1, y=diff_1, col sep=comma] {DataForPlots.csv};
\addplot table [color=red, x=time_5, y=diff_5, col sep=comma] {DataForPlots.csv};
\addplot table [color=green, x=time_10, y=diff_10, col sep=comma] {DataForPlots.csv};
\addplot table [color=black, x=time_25, y=diff_25, col sep=comma] {DataForPlots.csv};
\addplot table [color=brown, x=time_50, y=diff_50, col sep=comma] {DataForPlots.csv};
    \legend{1 batch, 5 batches, 10 batches, 25 batches, 50 batches}
\end{axis}
\end{tikzpicture}
\begin{tikzpicture}[yscale=0.9, xscale=0.8]
\begin{axis}[
    xlabel={Time(secs)},
    ylabel={Bounds Differences},
    xmin=0, xmax=1020,
    ymin=0, ymax=5,
    ytick={0, 2.5, 5},
    legend pos=north east
]

\addplot table [x=time_r_5d, y=diff_r_5d, col sep=comma] {DataForPlots.csv};
\addplot table [x=time_25, y=diff_25, col sep=comma] {DataForPlots.csv};
\addplot table [x=time_multi_5d, y=diff_multi_5d, col sep=comma] {DataForPlots.csv};
\legend{Method~II, Method~II-V5, Method~I}
\end{axis}
\end{tikzpicture}

\caption{Left: changes in results from different numbers of batches used among updates of the stopping strategy when we generate fresh data for training. Right: changes of the results using different methods: blue line: original method~II; red line: method~II with variation~5 and 25 batches were used among updates; brown line: method~I with variation~1. The top and bottom two plots correspond to the 1D put option and the 5D max-call option, respectively.}
\label{fig: var_fresh}
\end{figure}
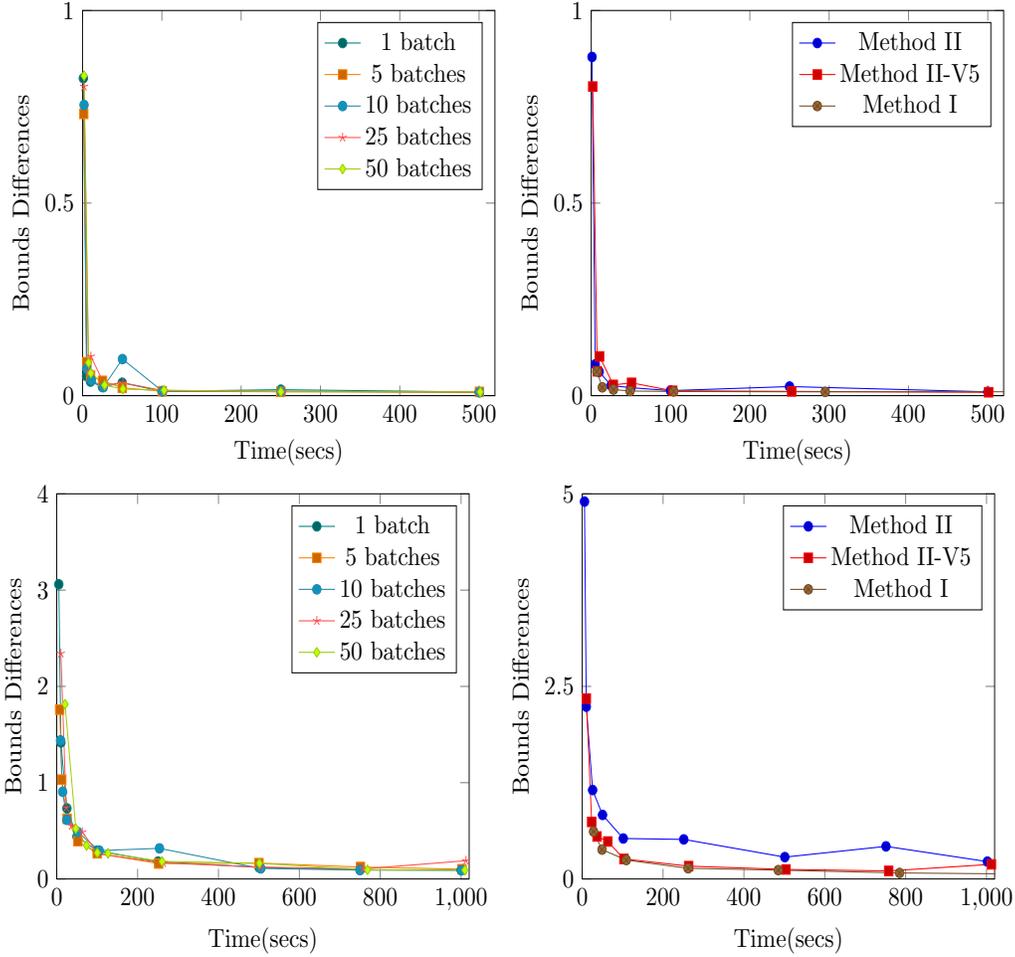

Figure \ref{fig: var_fresh} shows the difference between the lower and the upper bound of the option price throughout the training process for both the 1D put option (upper two plots) and the 5D max-call option (bottom two plots). The left two plots show these differences when employing various numbers of batches among updates. Initially, a higher number of batches leads to more favourable results. The difference diminishes when we train the model for a longer time. However, there is no definite conclusion on the optimal number of batches. 
The right two plots illuminate the difference using Method~I, Method~II, and Method~II with variation 5 and $25$ batches among updates. In all three cases, the second martingale term and separate networks are applied. All three schemes produce satisfactory results in pricing the 1D put option, but the base of Method~II performs worse when pricing the 5D max-call option. We can see that Method~I is more stable and converges faster among all three schemes, variant 5 exhibits superior performance to the base version.

\clearpage

\subsection*{Discussion}
We summarise the contributions of each variation brings to our methods in Table \ref{tab: VariationSummary} based on numerous experiments that were conducted apart from the ones demonstrated in this section, so the conclusion is generic. The second column indicates to which method one variation can be applied. There are three aspects one variation can affect: the accuracy of estimates, the training time and the computational memory required. We use {\cmark} and {\xmark} to indicate an improvement and a deterioration respectively. If a variation does not significantly impact one of these aspects, the corresponding cell remains blank.

From Table \ref{tab: VariationSummary}, we can see that variant 1 and variant 3 can improve the accuracy without prolonging the training time. Variant 2 improves the accuracy at the expense of the computational speed, while the opposite is true for variant 4. Variation 5 is the only one that helps us overcome the memory exhaustion problem, and compared to the base of Method~II, it also produces narrower bounds differences with shorter running time. All variations that can be applied to one method can be used at the same time to combine their effects.

In addition to these five variants, we also tested a warm-start approach applied to Method~I, where previously trained network is used as the initial network at the next step. Since it is a standard method in the field, its impact is detailed in Appendix \ref{app: warm-start}.

\begin{table}[htbp]
  \centering
    \begin{tabular}{|l|c|c|c|c|}
    \hline
     Variations & Method &Accuracy &Time &Memory \\
    \hline
      V1: Add a second martingale term&I, II & \cmark & & \\
    \hline
      V2: Add sub-steps&I, II & \cmark & \xmark & \\
    \hline
      V3: Use two separate networks&I, II & \cmark &  & \\
    \hline
      V4: Train on partial data&II & \xmark & \cmark & \\
    \hline
      V5: Train on fresh data&II & \cmark & \cmark & \cmark\\
    \hline
    \end{tabular}%
  
  \caption{Algorithm variants and their impacts on three different aspects. \cmark~and \xmark~represent an improvement and deterioration, respectively.}
  \label{tab: VariationSummary}%
\end{table}%

\section{Numerical results} \label{sec: numerical_results}
This section presents the numerical results obtained through both methods we proposed, incorporating variations 1 and 3 in Method~I, and variations 1, 3, and 5 in Method~II. The warm-start training has also been applied to Method~I. Our experimental setup includes the use of the ADAM optimizer, and mean squared error for the loss function. Softplus is chosen as the activation function due to its smoothness property. We standardise all input variables, except for the time variable in Method~II. To mitigate overfitting, cross-validation is rigorously applied throughout the training phase. The out-of-sample test set has $10^{6}$ paths across all scenarios. Networks with different structures were used in different cases, as detailed in each subsection. The selection of a specific network is based on extensive experimentation. We choose the ones that require the least hyperparameter tuning. Computations were executed on an NVIDIA P100 GPU under the system Intel Xeon-E5-2680-v4. The program is written in Python 3.8.5 using PyTorch 1.8.

Subsequent subsections demonstrate statistics of the pricing results for each option priced by repeating the process $10$ times, including means and standard deviations of the lower bound, the upper bound and their difference. The total running time (in seconds) for each repetition is also recorded.

Additionally, we plot histograms to depict both the total hedging errors $\epsilon_1$ and the worst hedging errors $\epsilon_2$ using an independent dataset containing $10^6$ paths. These metrics are computed as follows. Let $\tau^{i}$ be the stopping time for path $i$. The error for that path at $\tau^i$ is defined as $$\epsilon^i_1 = V_0+\sum_{t=t_0}^{\tau^i-\Delta t} \beta_{t}^{-1}H(S_t^i)\Delta W_t^i-\beta_{\tau^i}^{-1}Z^i_{\tau^i},$$ and the worst error is defined as
$$\epsilon^i_2=V_0+\min_{t^i \in \{t_1, ..., t_n\}}\!\left(\sum_{t=t_0}^{t^i-\Delta t} \beta_{t}^{-1}H(S^i_t)\Delta W^i_t-\beta_{t^i}^{-1}Z^i_{t^i}\right).$$

\subsection{Options under the Black--Scholes model}

Consider American-style options with $d$ underlying assets, whose prices follow the dynamics
$$
\D{S_{t}^{i}} = (r - \delta^{i}) S_{t}^{i} \D{t} + \sigma^{i} S_{t}^{i} \D{W^{i}_{t}},
$$
for $i \in \{1, 2, \ldots, d\}$, where the risk-free interest rate $r\in\mathbb{R}$, the dividend rate $\delta^i\in\RR$ and the volatility $\sigma^i\in\mathbb{R}^+$. Each Brownian Motion is independent of the others.

\subsubsection{1D American-style put option\label{ref:1D-American_put-option}}
We first test our method on a 1D vanilla American-style put option with the following parameters:
$$ T=1,\ K=40,\ n=50,\ S_{0}=36,\ r=0.06,\ \delta=0,\ \sigma=0.2$$ 
where $T$ is the maturity, $K$ is the strike price, and $n$ is the number of exercise opportunities. We use the same notation for all cases in this section. The payoff function at $t$ is $$f(S_t, K)=(K-S_t)^+.$$ We use $10^5$ paths to train the model in Method~I, and $7.63 \times 10^{6}$ in Method~II. The difference in the number of paths used is caused by the nature of method~II, which stops when the learning stagnates. This means that the number $7.63 \times 10^{6}$ is only an upper bound on the actual number of paths used. Moreover, method~II is designed to not remember any path, so that there is no challenge on the memory budget.

The benchmark computed by the finite difference method is $4.478$. The results generated by our schemes are shown in Table \ref{tab: 1DBS}. In method~I, at each time the training ceases once the loss of the validation set stagnates for $20$ epochs. In method~II, the training stops when the validation set loss stagnates for more than $5$ updates, and we train $20$ batches for $20$ epochs among updates. Method~I uses two networks with the structures ([1, 20, 20, 1],[1, 20, 20, 2]) at each time, while Method~II employs a total of two networks with structures ([2, 20, 20, 20, 1], [2, 20, 20, 20, 2]). The total numbers of parameters trained are $49150$ and $1863$ for Methods I and II, respectively. From Table \ref{tab: 1DBS}, we can see that both methods generate tight bounds. Even though Method~I exhibits shorter training times, it involves significantly more free variables in the training process. 

\begin{figure}[ht]
    \centering
    \includegraphics[width=0.5\paperwidth]{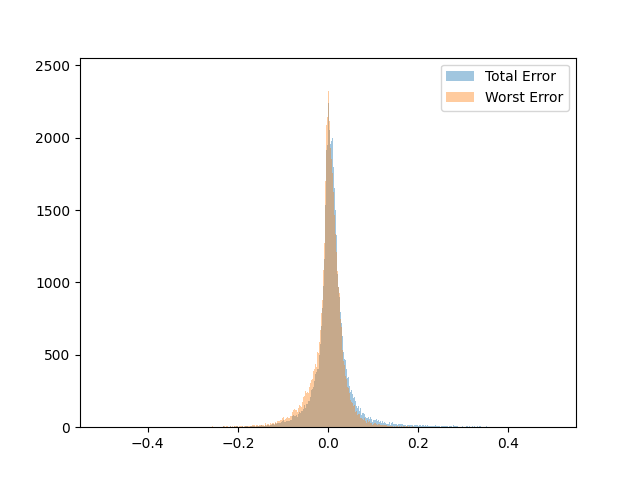}
    \caption{Hedging errors for the 1D American-style put option along 100,000 paths by directly using the model we trained via method~I.}
    \label{fig: Hedging1D}
\end{figure}

Figure \ref{fig: Hedging1D} shows the hedging errors defined at the start of this section. We can see that both hedging errors are distributed close to zero and the shape is symmetric. Their means, standard deviations and the ratio of the standard deviation to the estimated option value are shown in Table \ref{tab: 1DHedging}.

\begin{table}[htbp]
  \centering
    \begin{tabular}{|c|c|c|c|}
    \hline
     & Mean & S.D. & Mean/$\Hat{V}$\\
    \hline
    Total Error & $6.7004 \times 10^{-7}$ & 0.05165 & $1.4965 \times 10^{-7}$ \\
    \hline
    Worst Error & $-1.1479 \times10^{-2}$ & 0.04367 & $-2.564 \times10^{-3}$\\
    \hline
    \end{tabular}%
    \caption{Hedging errors for the 1D American-style put option with 100,000 paths using the model trained via method~I.}
  \label{tab: 1DHedging}%
\end{table}%

\begin{table}[htbp]
  \centering
    \begin{tabular}{|c|c|c|c|c|c|c|c|}
    \hline
        \multicolumn{2}{|c|}{} & \multicolumn{2}{c|}{LB} & \multicolumn{2}{c|}{UB} & \multicolumn{2}{c|}{Diff} \\
    \hline
     & Time & Mean & S.D. & Mean & S.D. & Mean & S.D. \\
    \hline
    I & 70 &4.4770 & 0.0003 & 4.4899 & 0.0006 & 0.0129 & 0.0008\\
    \hline
    II & 93 & 4.4749 & 0.0008 & 4.4880 & 0.0008 & 0.0131 &0.0014\\
    \hline
    \end{tabular}%
    \caption{1D American-style put option pricing using both schemes. Benchmark estimate: $4.478$. The first column indicates the method used. Method~I uses two networks with structures ([1, 20, 20, 1],[1, 20, 20, 2]) at each time, while Method~II uses in total two networks with structures ([2, 20, 20, 20, 1], [2, 20, 20, 20, 2]).}
  \label{tab: 1DBS}%
\end{table}%

\subsubsection{High-dimensional Bermudan max-call option}
Consider an option with $d$ underlying assets. We assume there is no correlation between Brownian Motions $W^i$ and $W^j$, $i, j\in\{1, 2, ..., d\}$, on which each stock price is based. The model has the following parameters:
$$T=3,\ K=100,\ n=9,\ r=0.05,\ \delta^{i}=0.1,\ \sigma=0.2.$$
The payoff of this option is 
$$\left(\max_{i\in\{1, 2, ..., d\}} S_{t}^{i} - K\right)^{+}.$$  
Given the sizeable step interval of ${\Delta}t=\frac{1}{3}$, we engage variation 2, employing $32$ substeps in our experiments. 
In Method~I, training concludes when the validation set's loss ceases to decrease after $5$ epochs. In method~II, the training stops when the validation set loss stagnates for more than $5$ updates, and we train $20$ batches for $20$ epochs among updates. Method~I consistently employs $10^{6}$ training paths, while Method~II's path count varies due to the nature of the scheme, as detailed in Table \ref{tab: max-call_paths}. 
\begin{table}[htbp]
  \centering
    \begin{tabular}{|c|c|c|c|}
    \hline
        $d$ & $S_{0}=90$ & $S_{0}=100$ & $S_{0}=110$   \\
    \hline
       5 & $1.189 \times10^{7}$ & $1.083\times10^{7}$ & $1.239\times10^{7}$ \\
    \hline
       10 & $7.98 \times10^{6}$ & $1.056\times10^{7}$ & $9.4\times10^{6}$\\
    \hline
    \end{tabular}%
    \caption{The number of training paths used in pricing the Bermudan max-call option using method~II. The first column shows the number of underlying assets.}
  \label{tab: max-call_paths}%
\end{table}%

\begin{figure}[ht]
    \centering
    \includegraphics[width=0.5\paperwidth]{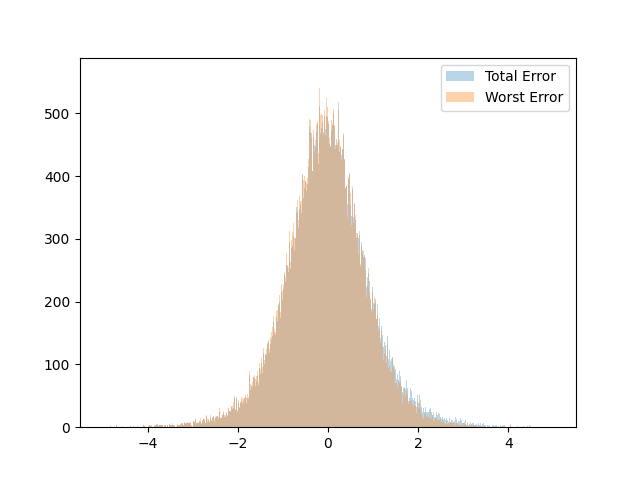}
    \caption{Hedging errors of 5D Bermudan max-call option with 32 sub-steps.}
    \label{fig: hedging5D}
\end{figure}

Table \ref{tab: max-call} presents the pricing results of a max-call option with three different initial stock prices in both 5D and 10D settings. The benchmark given is extracted from \citet{becker2020pricing}, including the approximated bounds where the left (right) value is the lower (upper) bound (the number on the top), alongside the aggregate computation time (the summation at the bottom). The numbers in the summation are calculation time in seconds for lower bounds, upper bounds and hedging strategies, respectively. The benchmark duration for hedging reflects the time to formulate a complete hedging strategy from $0$ to $T$ with $96$ substeps, chosen due to the resemblance of its hedging error to our results, as illustrated in Figure \ref{fig: hedging5D} and  Table \ref{tab: 5DHedging}.

\begin{table}[htbp]
  \centering
    \begin{tabular}{|c|c|c|c|}
    \hline
     & Mean & S.D. & S.D./$\Hat{V}$ \\
    \hline
    Total Error & $\hspace{0.75em}1.0492 \times 10^{-6}$ & 0.9612 & $\hspace{0.75em}4.012 \times 10^{-8}$\\
    \hline
    Worst Error & $-6.9951 \times 10^{-2}$ & 0.9616 & $-2.675 \times 10^{-3}$\\
    \hline
    \end{tabular}%
    \caption{Hedging errors of the 5D max-call option where $S_0=100$ with 100,000 paths using the model trained via method~I.}
  \label{tab: 5DHedging}%
\end{table}%

From Table \ref{tab: max-call}, we can see that the empirical results from both methods align closely with the benchmarks yet achieving quicker total computation times. The gaps between lower and upper bounds in the benchmark are relatively tighter than the ones derived from our methods. This is because options being priced have infrequent exercise times which allows them to use nested Monte Carlo to derive more accurate results, while our method is designed to derive both bounds and hedging strategies simultaneously with lower computational cost. From the table we can see that our lower bounds invariably remain beneath our upper bounds; by contrast, there is one contradicting case in their results. In addition, their method is faster in deriving the price bounds but slower in deriving hedging strategies comparing to ours. However, we acknowledge that the difference in computational time can be partially contributed by different computing environments and other programming related factors. Our method also generates smaller hedging errors on average. We can also see that our methods become more competitive in the 10D case, and this is particularly true for Method~II which can generate tighter bounds with less running time. This advantage is expected to become increasingly significant in more complicated cases, given its lower requirement for computing resources. 

\begin{table}[ht]
  \centering
    \begin{tabular}{|c|c|c|c|c|c|c|c|c|c|c|}
    \hline
        \multicolumn{4}{|c|}{} & \multicolumn{2}{|c|}{LB} & \multicolumn{2}{c|}{UB} & \multicolumn{2}{c|}{Diff} &  \\
    \hline
       \multicolumn{1}{|c|}{$d$} & \multicolumn{1}{c|}{$S_0$} &  & \multicolumn{1}{c|}{Time} & \multicolumn{1}{c|}{Mean} & \multicolumn{1}{c|}{S.D.} & \multicolumn{1}{c|}{Mean} & \multicolumn{1}{c|}{S.D.} & \multicolumn{1}{c|}{Mean} & \multicolumn{1}{c|}{S.D.} & \multicolumn{1}{c|}{Benchmark} \\
    \hline
       \multirow{6}[0]{*}{5} & \multirow{2}[0]{*}{90} & I     & 1002  & 16.6377 & 0.0009 & 16.6862 & 0.0023 & 0.0484 & 0.0026 & (16.644, 16.648) \\
    \cline{3-10}
          &       & II    &  1127 &16.6314&0.0030&16.6856 & 0.0025 & 0.0542 & 0.0040      &   132+8+1546\\
    \cline{2-11}
          & \multirow{2}[0]{*}{100} & I     & 1022  & 26.1523 & 0.0012 & 26.2195 & 0.0021 & 0.0672 & 0.0029 & (26.156, 26.152) \\
    \cline{3-10}
          &       & II    & 1034  & 26.1411 & 0.0050 & 26.2259 & 0.0029 & 0.0848 & 0.0067 &  134+8+1668\\
    \cline{2-11}
          & \multirow{2}[0]{*}{110} & I &  1177 & 36.7551 & 0.0252 & 36.8724 & 0.0068 & 0.1173 & 0.0311 & (36.780, 36.796) \\
    \cline{3-10}
          &       & II    & 1038  & 36.7767 & 0.0013 & 36.8646 & 0.0024 & 0.0879 & 0.0024 &  133+8+1673\\
    \hline
    \multirow{6}[0]{*}{10} & \multirow{2}[0]{*}{90} & I     & 989   & 26.2613 & 0.0057 & 26.4823 & 0.0177 & 0.2210 & 0.0226 & (26.277, 26.283) \\
     \cline{3-10}
          &       & II    & 788   & 26.2446 & 0.0200  & 26.3822 & 0.0162 & 0.1376 & 0.0353 &  136+8+1792\\
   \cline{2-11}
          & \multirow{2}[0]{*}{100} & I     & 1035  & 38.3503 & 0.0067 & 38.5974 & 0.0286 & 0.2471 & 0.0345 & (38.355, 38.378) \\
     \cline{3-10}
          &       & II    & 1045 & 38.3159 & 0.0235 & 38.4894 & 0.0144 & 0.1735 & 0.0375 & 136+7+1803 \\
     \cline{2-11}
          & \multirow{2}[0]{*}{110} & I     & 1023  & 50.8961 & 0.0047 & 51.1810 & 0.0184 & 0.2849 & 0.0228 & (50.869, 50.932) \\
     \cline{3-10}
          &       & II    & 928   & 50.8764 & 0.0151 & 51.0601 & 0.0083 & 0.1837 & 0.0197 &  135+8+1777\\
    \hline
    \end{tabular}%
    \caption{Bermudan max-call option pricing using both methods. The first and the second columns show the number of underlying assets and the initial stock price, respectively. The third column indicates the method used. Benchmarks in the last column are extracted from \citep{becker2020pricing}. The top value is the estimated price bounds, the three values shown below are the times for deriving a lower bound, an upper bound and a hedging strategy. Method~I uses two networks with structures ($[d, 50, 25, 1]$, $[d, 50, 50, 2d]$) at each time, while method~II uses in total two networks with structures ($[d+1, 50, 50, 50, 1]$, $[d+1, 50, 50, 50, 2d]$).}
  \label{tab: max-call}%
\end{table}%

\subsubsection{High dimensional American-style geometric-put option}
Consider a geometric option with $d$ underlying assets, where all stocks have the same dynamics. The parameters used for the stocks dynamics and the option payoff are:
$$S_{0}^{i}=1, \quad r=0.05, \quad \sigma=0.2, \quad T=1, \quad K=1.$$ 
The payoff of this option is 
$$f(S_{t})=\left(K - \prod_{j = 1}^{d}S_{t}^{j}\right)^{+}.$$ 
Scenarios with different numbers of stocks and exercise opportunities are tested. Due to the property of the geometric payoff, this option with multiple underlying assets can be valued by a 1D put option by adjusting the parameters of the underlying:
$$ \hat{S_{0}} = \prod_{i=1}^{d}S_{0}^{i}, \quad 
   \hat{r} = r\cdot d, \quad
   \hat{\sigma} = \sqrt{d} \cdot \sigma.$$
   
\begin{table}[ht]
  \centering
    \begin{tabular}{|c|c|c|c|c|c|c|c|c|c|c|}
    \hline
        \multicolumn{4}{|c|}{} & \multicolumn{2}{c|}{LB} & \multicolumn{2}{c|}{UB} & \multicolumn{2}{c|}{Diff} & \multicolumn{1}{c|}{Benchmark} \\
    \cline{1-10}
       \multicolumn{1}{|c|}{$d$} & \multicolumn{1}{c|}{$n$} &  & \multicolumn{1}{c|}{Time} & \multicolumn{1}{c|}{Mean} & \multicolumn{1}{c|}{S.D.} & \multicolumn{1}{c|}{Mean} & \multicolumn{1}{c|}{S.D.} & \multicolumn{1}{c|}{Mean} & \multicolumn{1}{c|}{S.D.} & \multicolumn{1}{c|}{(BT)} \\
    \hline
       \multirow{8}[0]{*}{5} & \multirow{2}[0]{*}{10} & I     & 
    235   & 0.1049 & 0.0001 & 0.1129 & 0.0002 & 0.0079 & 0.0002 &  \multirow{8}[0]{*}{0.1072}\\
    \cline{3-10}
     &  & II & 219   & 0.1041 & 0.0002 & 0.1152 & 0.0016 & 0.0111 & 0.0016 &  \\
     \cline{2-10}
    & \multirow{2}[0]{*}{20}  & I & 402   & 0.1056 & 0.0001 & 0.1173 & 0.0015 & 0.0116 & 0.0016 &  \\
    \cline{3-10}
    & & II &
    375   & 0.1047 & 0.0006 & 0.1201 & 0.0038 & 0.0154 & 0.0038 &  \\
    \cline{2-10}
    & \multirow{2}[0]{*}{40} & I &
    701   & 0.1062 & 0.0001 & 0.1172 & 0.0018 & 0.011 & 0.0019 &  \\
    \cline{3-10}
    & & II &
    691   & 0.1045 & 0.0004 & 0.1243 & 0.0027 & 0.0199 & 0.0028 &  \\
    \cline{2-10}
     & \multirow{2}[0]{*}{80}& I &
    1302  & 0.1067 & 0.0001 & 0.1153 & 0.0013 & 0.0086 & 0.0014 &  \\
    \cline{3-10}
    & & II &
    1430  & 0.1044 & 0.0007 & 0.1277 & 0.0009 & 0.0234 & 0.0012 &  \\
    \hline
    \multirow{8}[0]{*}{10} & \multirow{2}[0]{*}{10} & I & 269   & 0.1256 & 0.0001 & 0.1396 & 0.002 & 0.014 & 0.0021 &  \multirow{8}[0]{*}{0.1296}\\
     \cline{3-10}
          &    &  II  & 188   & 0.1241 & 0.0003 & 0.1423 & 0.0033 & 0.0182 & 0.0035 &  \\
           \cline{2-10}
          & \multirow{2}[0]{*}{20} & I & 466   & 0.1276 & 0.0001 & 0.1389 & 0.0004 & 0.0113 & 0.0005 &  \\
           \cline{3-10}
          &   & II     & 422   & 0.1251 & 0.0003 & 0.1464 & 0.0034 & 0.0214 & 0.0036 &  \\
           \cline{2-10}
          & \multirow{2}[0]{*}{40} & I & 757   & 0.1284 & 0.0001 & 0.1406 & 0.0006 & 0.0122 & 0.0007 &  \\
           \cline{3-10}
          &   & II     & 856   & 0.1252 & 0.001 & 0.1506 & 0.0038 & 0.0254 & 0.0043 &  \\
           \cline{2-10}
          & \multirow{2}[0]{*}{80} & I & 1399  & 0.1288 & 0.0001 & 0.1434 & 0.0021 & 0.0145 & 0.0022 &  \\
           \cline{3-10}
          &  & II   & 1364  & 0.1236 & 0.0011 & 0.1553 & 0.0018 & 0.0317 & 0.0023 &  \\
          \hline
    \multirow{8}[0]{*}{20} & \multirow{2}[0]{*}{10} & I & 314   & 0.1437 & 0.0001 & 0.1599 & 0.0003 & 0.0162 & 0.0003 &  \multirow{8}[0]{*}{0.1502}\\
     \cline{3-10}
          &   & II    & 217   & 0.1412 & 0.0009 & 0.1658 & 0.0028 & 0.0247 & 0.0028 &  \\
           \cline{2-10}
          & \multirow{2}[0]{*}{20} & I & 509   & 0.1468 & 0.0001 & 0.1646 & 0.0009 & 0.0177 & 0.0009 &  \\
           \cline{3-10}
          &   & II    & 472   & 0.143 & 0.0015 & 0.1777 & 0.0147 & 0.0347 & 0.016 &  \\
           \cline{2-10}
          & \multirow{2}[0]{*}{40} & I & 869   & 0.1484 & 0.0001 & 0.1692 & 0.0023 & 0.0208 & 0.0024 &  \\
           \cline{3-10}
          &  &  II   & 660   & 0.143 & 0.0019 & 0.2061 & 0.0201 & 0.0631 & 0.0212 &  \\
            \cline{2-10}
          & \multirow{2}[0]{*}{80} & I & 1341  & 0.1488 & 0.0002 & 0.1769 & 0.0021 & 0.0281 & 0.0023 &  \\
           \cline{3-10}
          &   & II    & 1082  & 0.1424 & 0.0021 & 0.2213 & 0.026 & 0.0789 & 0.026 &  \\
           \hline
    \end{tabular}%
    \caption{Pricing geometric put option using both methods. The first and the second columns show the number of underlying assets and the number of exercise points, respectively. The third column indicates the method used. Benchmarks in the last column are derived by the binomial tree method. Method~I uses two networks with structures ($[d, 50, 50, 1]$, $[d, 50, 50, 2d]$) at each time, while method~II uses in total two networks with structures ($[d+1, 20, 20, 20, 1]$, $[d+1, 20, 20, 20, 2d]$).}
  \label{tab: geo-put}%
\end{table}%
   
Table \ref{tab: geo-put} shows the numerical results generated by our methods. They are dedicated to options with $d \in \{5, 10, 20\}$ underlying assets. For each option, $n=10, 20, 40, 80$ steps have been used to price the option. The reference values are calculated by the binomial tree method. We can see that the results derived by the proposed methods are close to the benchmark, showing their capability to price high-dimensional options with frequent exercises.


\clearpage

\subsection{American-style put option under the Heston model}
Finally, we test our methods under the Heston model, where the volatility itself is also stochastic:
$$
\left\{\begin{array}{rcl}
    \D{S_{t}} &=& r S_{t} \D{t} + \sqrt{V_{t}} S_{t} \D{W_{S}}, \\
    \D{V_{t}} &=& \lambda (\sigma^{2} - V_{t}) \D{t} + \xi \sqrt{V_{t}} \D{W_{V}}.
\end{array}\right.
$$
The option we price is the same as the one in \citet{lapeyre2021}, characterized by the parameters: 
$$T=1, K=100, n = 10, S_{0}=100, V_{0}=0.01, r=0.1, \sigma=0.1, \lambda=2, \xi=0.2, \rho=-0.3,$$ 
where $\rho$ is the correlation between Brownian Motions $W_S$ and $W_V$.

Since there are two Brownian motions involved in this scenario and we apply variation~1 for enhanced precision, we have 
$$ \Psi^{S}_{1}(S_{t_{i}})\Delta W^S_{t_{i}}
  +\Psi^{V}_{1}(S_{t_{i}})\Delta W^V_{t_{i}}
  +\Psi^{S}_{2}(S_{t_{i}})\Delta ((W^S_{t_{i}})^2-\Delta t)
  +\Psi^{V}_{2}(S_{t_{i}})\Delta ((W^V_{t_{i}})^2-\Delta t)
$$ 
as our martingale increment. Similar to the max-call option in section $5.1.2$, the step size $\Delta t=0.1$ is big, so we use $9$ substeps for the implementation. 

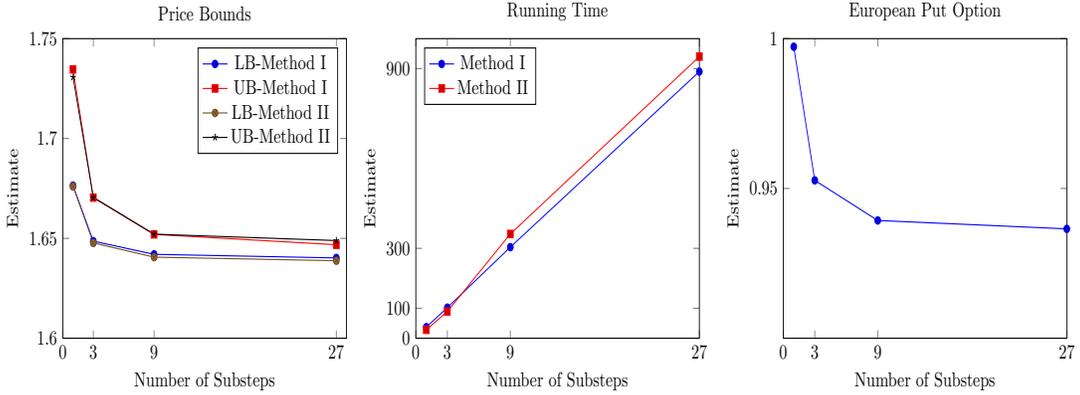
\begin{figure}[ht]
\centering
\begin{tikzpicture}[yscale=0.7, xscale=0.55]
\begin{axis}[
    title={Price Bounds},
    xlabel={Number of Substeps},
    ylabel={Estimate},xmin=0, xmax=28, ymin=1.6, ymax=1.75,
    xtick={0, 3, 9, 27},
    ytick={1.6, 1.65, 1.7, 1.75},
    legend pos=north east
]
\addplot table [x=sub_step_H_mul, y=lb_H_mul, col sep=comma] {DataForPlots.csv};
\addplot table [x=sub_step_H_mul, y=ub_H_mul, col sep=comma] {DataForPlots.csv};
\addplot table [x=sub_step_H_mul, y=lb_H_one, col sep=comma] {DataForPlots.csv};
\addplot table [x=sub_step_H_mul, y=ub_H_one, col sep=comma] {DataForPlots.csv};
\legend{LB-Method~I, UB-Method~I, LB-Method~II, UB-Method~II}
    
\end{axis}
\end{tikzpicture}
\begin{tikzpicture}[yscale=0.7, xscale=0.55]
\begin{axis}[
    title={Running Time},
    xlabel={Number of Substeps},
    ylabel={Estimate},xmin=0, xmax=27, ymin=0, ymax=1000,
    xtick={0, 3, 9, 27},
    ytick={0, 100, 300, 900},
    legend pos=north west
]
\addplot table [x=sub_step_H_mul, y=time_H_mul, col sep=comma] {DataForPlots.csv};
\addplot table [x=sub_step_H_mul, y=time_H_one, col sep=comma] {DataForPlots.csv};
\legend{Method~I, Method~II}
    
\end{axis}
\end{tikzpicture}
\begin{tikzpicture}[yscale=0.7, xscale=0.55]
\begin{axis}[
    title={European Put Option},
    xlabel={Number of Substeps},
    ylabel={Estimate},xmin=0, xmax=27, ymin=0.9, ymax=1,
    xtick={0, 3, 9, 27},
    ytick={09, 0.95, 1},
    legend pos=north west
]
\addplot table [x=sub_step_H_mul, y=euro_H, col sep=comma] {DataForPlots.csv};
    
\end{axis}
\end{tikzpicture}
\caption{Price bounds (\textit{Left}) and corresponding running times (\textit{Right}) of the Heston put option with different numbers of substeps using both method~I and II (variant~5).}
\label{fig: Heston_sub}
\end{figure}

Figure \ref{fig: Heston_sub} shows the change in the estimates with an increasing number of substeps using both methods. We can see that both lower and upper bounds decrease and the gap becomes narrower with decreasing step size due to the more accurate martingale increment approximation. However, as the estimated bounds decrease, the computational time rises.  

Given the insights from Figure \ref{fig: Heston_sub}, we choose to add $9$ substeps to approximate the option price, since further improvement becomes trivial while significantly increasing computational expenses, shown in Table \ref{tab: 2DHeston}. The estimated bounds we generated are tight, and the computation time can be very small with the adjustment of the training path. However, we can see that the lower bound also decreases which is opposite to the examples before. This is caused by the truncation error occurred when simulating the
Heston model using the Euler scheme. When the step size decreases, the model is better approximated due to the reduced bias from the truncation. The similar effect can be observed in European option pricing under the Heston model, shown in the last plot in Figure \ref{fig: Heston_sub}. Note that the total number of steps used in the simulation for the European option is the number of substeps times the $n$. In this table, we can see our results exhibit slight deviations from the benchmark. This is potentially attributable to model simulation variations. 

\begin{table}[htbp]
  \centering
    \begin{tabular}{|c|c|c|c|c|c|c|c|c|}
    \hline
        \multicolumn{3}{|c|}{} & \multicolumn{2}{c|}{LB} & \multicolumn{2}{c|}{UB} & \multicolumn{2}{c|}{Diff} \\
    \hline
          & \multicolumn{1}{|c|}{Path} & \multicolumn{1}{c|}{Time} & \multicolumn{1}{c|}{Mean} & \multicolumn{1}{c|}{S.D.} & \multicolumn{1}{c|}{Mean} & \multicolumn{1}{c|}{S.D.} & \multicolumn{1}{c|}{Mean} & \multicolumn{1}{c|}{S.D.} \\
    \hline
    \multirow{2}[0]{*}{I} & $1\times10^{5}$ & 36    & 1.6416 & 0.0001 & 1.6659 & 0.0033 & 0.0244 & 0.0033 \\
    \cline{2-9}
          & $1\times10^{6}$ & 294   & 1.6419 & 0.0001 & 1.6514 & 0.0005 & 0.0094 & 0.0005 \\
    \hline
    \multirow{2}[0]{*}{II} &
    $9.45\times10^{5}$     & 28    & 1.6364 & 0.0030 & 1.6979 & 0.0271 & 0.0615 & 0.0291 \\
    \cline{2-9}
          & $1.72\times10^{7}$ & 348& 1.6406 & 0.0007 &	1.6521	& 0.0007	& 0.0115& 	0.0008
 \\
    \hline
    \end{tabular}%
    \caption{1D American-style put option under the Heston model with $9$ substeps added. The benchmark from \citep{lapeyre2021} is $1.7\pm0.0016$. The second column indicates the number of training paths employed. Method~I uses two networks with structures ($[2, 50, 50, 1]$, $[2, 50, 50, 4]$) at each time, while method~II uses in total two networks with structures ($[3, 50, 50, 50, 1]$, $[3, 50, 50, 50, 4]$).}
  \label{tab: 2DHeston}%
\end{table}%

\section{Conclusion}
In this paper, we introduce two innovative approaches aimed at simultaneously addressing the American-style option pricing problem and its dual form, providing both lower and upper bounds on the option price using deep learning using neural networks. Both methods are based on the least squares Monte Carlo method with the incorporation of duality. The first method employs a series of networks to approximate the continuation values and martingale increments at each exercise time. The second method applies one global network by adding time as a state variable to perform the regression and alternates the network training and the update of the stopping strategy till a stopping criterion is met. We propose several variants to enhance the methods from different perspectives. One notable advantage of our methods is that nested simulations are avoided, significantly reducing the computation complexity when pricing American-style/Bermudan options that have frequent exercise opportunities. Moreover, the methods naturally yield hedging strategies, serving as effective control variates for variance reduction.  

Although the numerical results predominantly rely on the geometric Brownian Motion, it is important to emphasize that the applicability of our methods extends beyond this model. Our methods can take any model that can be simulated and satisfy conditions of the martingale representation theorem such that martingale increments can be approximated. The demonstrated effectiveness in pricing options within the Heston model underscores the versatility of our approaches. This property of our methods provides a ground for exploration, encouraging their application to problems in more complicated models. 

From the results shown, we can see that both methods yield tight bounds for the approximated option price in both low and high-dimensional cases. Though the training process can be time-intensive for high-dimensional problems, the resulting models can be directly used to derive a hedging strategy without additional effort. In both methods, the inclusion of a second martingale increment term and the introduction of substeps for options with less frequent exercise points play important roles in improving the accuracy. In conclusion, Method~I demonstrates greater stability and yields narrower bounds differences. However, its performance diminishes as the complexity of the problem increases and the required number of training paths grows too large. On the other hand, Method~II with the application of variation 5, effectively overcomes these challenges. This is evidenced by its successful pricing of the 10D max-call option. Further exploration of this variant could be conducted to fully assess its capabilities and potential enhancements. 

\section*{Code availability}
\noindent
Our code is openly available at: \\
\url{https://github.com/JiahaoWu27/American-Option-Pricing.git}

\section*{Disclosure of interest}
\noindent No potential competing interest was reported by the authors.

\bibliography{bibliography} 

\appendix
\section{Variance reduction}\label{app: VR}
We have mentioned the process $H$ can be approximated by a function of the stock price due to the Markovian property of the stock processes. From now on, let $H_{t}=H(S_{t})$.
\begin{proposition}
    Given the option has not been exercised at $t\in[0, T)$. Let $\tau^{*} \in \Tau_{t}$ be the optimal stopping time. The martingale increment $\int_{t}^{\tau^{*}} H(S_{u}) \D W_{u}$ can be used as control variate to reduce variance.
\end{proposition}
\begin{proof}
 Since $V_{\tau^{*}}$ is $\FF_{T}$ measurable, we can apply martingale representation:
\begin{align}
V_{\tau^*} 
&= \EV{V_{\tau^{*}}} + \int_{0}^{\tau^{*}} H(S_{u})\D W_{u} + \int_{\tau^{*}}^{T} H(S_{u})\D W_{u} \label{eq: VR1}
\end{align}
By taking expectations on both sides of \eqref{eq: VR1} conditioned on $\FF_{\tau^{*}}$, we can get
\begin{align*}
\EV{V_{\tau^{*}}\Big|\FF_{\tau^{*}}}
&=\EV{V_{\tau^{*}}} +\int_{0}^{\tau^{*}} H(S_u)\D W_u 
  +\EV{\int_{\tau^{*}}^{T} H(S_u)\D W_u|\FF_{\tau^{*}}}
\end{align*}
Note that $\int_{\tau^{*}}^{T} H(S_{u})\D W_{u}=0$. This can be interpreted through the theory of hedging. We have $H_{t}=0$ for $t \in (\tau^{*}, T]$ because we stop hedging once the option is exercised. Hence, 
\begin{equation}\label{eq: VR2}
V_{\tau^{*}} 
= \EV{V_{\tau^{*}}} + \int_{0}^{t} H(S_u)\D W_{u}  
  + \int_{t}^{\tau^{*}} H(S_{u})\D W_{u}. 
\end{equation}
We then take expectations on both sides of \eqref{eq: VR2} conditioned on $\FF_{t}$:
\begin{align}
\EV{V_{\tau^{*}}|\FF_{t}} &= \EV{\EV{V_{\tau^{*}}}|\FF_{t}} + \EV{\int_{0}^{t} H(S_{u})\D W_{u}|\FF_{t}} +\EV{\int_{t}^{\tau^{*}} H(S_u)\D W_u|\FF_{t}}\nonumber\\
 &=\EV{V_{\tau^{*}}}+\int_{0}^{t} H(S_u)\D W_u \nonumber \\
 &= V_{\tau^{*}} - \int_{t}^{\tau^{*}} H(S_u)\D W_u. \label{eq: VR4} 
\end{align}
We can have
\begin{equation*}
 \EV{\EV{V_{\tau^{*}}|\FF_{t}} \cdot \int_{t}^{\tau^{*}} H(S_u)\D W_u\Bigg| \FF_{t}} 
 = \EV{V_{\tau^{*}}|\FF_{t}} \EV{\int_{t}^{\tau^{*}} H(S_u)\D W_u \Bigg| \FF_{t}} 
 = 0,    
\end{equation*}
and 
\begin{equation*}
 \EV{\EV{V_{\tau^{*}}|\FF_{t}}}\EV{\int_{t}^{\tau^{*}} H(S_u)\D W_u} = 0,    
\end{equation*}
implying $\EV{V_{\tau^{*}}|\FF_{t}}$ and $ \int_{t}^{\tau^{*}} H(S_u)\D W_u$ are uncorrelated given $\FF_{t}$, so
\begin{align} \label{eq: VR6}
\Var{V_{\tau^{*}}}
&=\Var{\EV{V_{\tau^{*}}|\FF_{t}}} + \Var{\int_{t}^{\tau^{*}} H(S_u)\D W_u}.
\end{align}
Combine \eqref{eq: VR4} and \eqref{eq: VR6}, we have
\small
\begin{equation} \label{eq: VRResult}
\Var{V_{\tau^{*}}-\int_{t}^{\tau^{*}} H(S_u)\D W_u} 
= \Var{V_{\tau^{*}}} - \Var{\int_{t}^{\tau^{*}} H(S_u)\D W_u} 
\leq \Var{V_{\tau^{*}}}.
\end{equation}
\normalsize
Hence, by subtracting the term $\int_{t}^{\tau^{*}} H(S_u)\D W_u$ from $V_{\tau^{*}}$, the variance is reduced.

\end{proof}
Therefore, adding the control variate $\int_{t_i}^{\tau_{i}} H_s dW_s$ in the derivation of $Y_{t_i}$ reduces the variance.

Note that we also show that $\int_{\tau_i}^{T} H_sdW_s=0$. This is in line with the hedging theory as we stop hedging once the stopping time is reached (the option is exercised).

\section{Warm-start training with the network trained one step before}\label{app: warm-start}
In the original version of method~I, no technique for parameter initialisation is employed, resulting in the random initialisation of weights and biases at the onset of the training. While this randomness typically does not pose a problem in practice, there is a possibility for training stagnation from the start due to subpar parameter choices, and it can lead to time-consuming processes. To enhance efficiency, we adopt a strategy where the parameters of a previously trained network serve as the initial values for the model under the current training. The rationale behind this technique is the observed similarities in the shapes of both continuation functions and martingale increment functions at different times, as shown in Figure \ref{fig: FuncsPlot}, suggesting that parameters across different networks should exhibit similarities. Table \ref{tab: var1_warm_start} and Figure \ref{fig: var1_warm_start} illustrate the impact of random and non-random initialisation on results. 

The table highlights that with this variant more accurate results are achievable in one-third of the time required for the base scheme. The enhancement in accuracy can be attributed to a better initial guess, facilitating more effective training in the right direction. The figure affirms the effectiveness of this modification. At the step before maturity, both versions commence with random initialisation, resulting in similar numbers of epochs. However, this number significantly decreases for all other steps. In most steps, less than half the number of epochs is needed. This effect is particularly pronounced at the initial time where the same $S_{0}$ is used for all paths. Although $\Delta W_{t_{0}}$ values differ, the training becomes highly versatile. In the present example, the number of epochs is $10$ times greater without the variant. This ratio can vary due to the randomness, with observed instances ranging from a worst-case scenario of $80$ times more to a best-case scenario of $3$ times more in experiments.

\begin{table}[htbp]
  \centering
    \begin{tabular}{|c|c|c|c|c|c|c|c|}
    \hline
    \multicolumn{2}{|c|}{} & \multicolumn{2}{|c|}{Lower Bound}       & \multicolumn{2}{|c|}{Upper Bound} &    \multicolumn{2}{|c|}{Difference}   \\
    \hline
      & Time(sec) & Mean &S.D. &Mean &S.D. & Mean & S.D. \\
    \hline
    Random start & 360 & 4.4738 & 0.0007 & 4.4889 & 0.0005 & 0.0151 & 0.0010   \\
    \hline
    Warm start & 135   & 4.4769 & 0.0002 & 4.4877 & 0.0004 & 0.0108 & 0.0005 \\
    \hline
    \end{tabular}%
  \caption{Pricing 1D vanilla American-style put option (with the same parameters as the ones in Section~\ref{ref:1D-American_put-option}). The first row displays results where weights are randomly initialised at each time. The second row shows the estimate when warm-start is applied.}
  \label{tab: var1_warm_start}%
\end{table}%
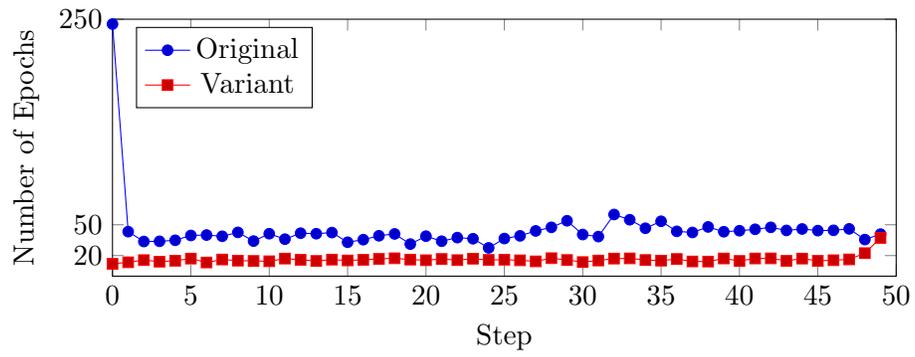
\begin{figure}[ht]
\centering
\begin{tikzpicture}[yscale=1.0, xscale=1.0]
\begin{axis}[
    height = 5cm,
    width = 12cm,
    xlabel={Step},
    ylabel={Number of Epochs},
    xmin=0, xmax=50,
    ymin=0, ymax=250,
    ytick={20, 50, 250},
    legend pos=north west
]

\addplot table [x=time_var1, y=random, col sep=comma] {DataForPlots.csv};
\addplot table [x=time_var1, y=non-random, col sep=comma] {DataForPlots.csv};
\legend{Original, Variant}  

\end{axis}
\end{tikzpicture}

\caption{The number of epochs needed till the training stagnates at different steps with/without this variant when pricing a 1D vanilla American put option.}
\label{fig: var1_warm_start}
\end{figure}

\end{document}